\documentclass[acmtog,authorversion,nonacm]{acmart}

\AtBeginDocument{%
  \providecommand\BibTeX{{%
    \normalfont B\kern-0.5em{\scshape i\kern-0.25em b}\kern-0.8em\TeX}}}

\setcopyright{acmcopyright}
\copyrightyear{2025}
\acmYear{2025}
\acmDOI{10.1145/3764928}

\acmJournal{TOG}
\acmVolume{44}
\acmNumber{4}
\acmArticle{}
\acmMonth{9}

\usepackage{xcolor}

\makeatletter
\newcommand{\ch}[1]{%
  \@bsphack
  \@esphack
  #1%
}
\makeatother

\makeatletter
\newcommand{\chf}[1]{%
  \@bsphack
  \@esphack
  #1%
}
\makeatother

\newcommand{\chn}[1]{#1}

\newif\iflong
\usepackage{amsmath}
\usepackage{dsfont}
\usepackage{subcaption}
\usepackage{multirow}
\usepackage{siunitx}
\usepackage{enumitem}
\usepackage[ruled, vlined, linesnumbered]{algorithm2e}
\usepackage[utf8]{inputenc}

\newcommand{\ForEachPar}[1]{%
  \ForEach{#1 \underline{\smash{\textbf{\textup{in parallel}}}}}%
}

\usepackage[final]{listings}
\usepackage[most]{tcolorbox}

\definecolor{rwthmagenta}{RGB}{200, 70, 158}
\definecolor{rwthblue}{rgb}{0,0.2,1.0}
\definecolor{typeblue}{RGB}{23, 110, 191}
\definecolor{backgroudblue}{rgb}{0.0, 0.5, 1.0}
\definecolor{codepurple}{rgb}{0.58, 0, 0.82}
\definecolor{codegreen}{rgb}{0, 0.6, 0}
\definecolor{codegray}{rgb}{0.61, 0.61, 0.61}
\colorlet{codebackcolor}{backgroudblue!3}

\newtcblisting{mycode}[1]{%
  boxsep=-5pt,
  boxrule=0.25pt,
  arc=2mm,
  auto outer arc,
  colframe=rwthblue,
  colback=codebackcolor,
  listing options={language=C++},
  listing only,
  after skip=0.4cm,
  #1
}

\newtcblisting{mycodescriptsize}[1]{%
  boxsep=-5pt,
  boxrule=0.25pt,
  arc=2mm,
  auto outer arc,
  colframe=rwthblue,
  colback=codebackcolor,
  listing options={language=C++, basicstyle=\linespread{0.90}\color{black!95}\ttfamily\scriptsize},
  listing only,
  after skip=0.4cm,
  #1
}

\newtcblisting{mycodetiny}[1]{%
  boxsep=-5pt,
  boxrule=0.25pt,
  arc=2mm,
  auto outer arc,
  colframe=rwthblue,
  colback=codebackcolor,
  listing options={language=C++, basicstyle=\linespread{0.90}\color{black!95}\ttfamily\tiny},
  listing only,
  after skip=0.4cm,
  #1
}

\newtcblisting{mycodeblock}[1]{%
 boxsep=-5pt,
 boxrule=0.25pt,
 arc=2mm,
 auto outer arc,
 colframe=rwthblue,
 colback=codebackcolor,
 listing options={language=C++,numbers=none, xleftmargin=-10pt},
 listing only,
 after skip=7pt,
 #1
}

\newtcblisting{mycodeblockscriptsize}[1]{%
 boxsep=-5pt,
 boxrule=0.25pt,
 arc=2mm,
 auto outer arc,
 colframe=rwthblue,
 colback=codebackcolor,
 listing options={language=C++,numbers=none, xleftmargin=-10pt, basicstyle=\linespread{0.90}\color{black!95}\ttfamily\scriptsize},
 listing only,
 after skip=7pt,
 #1
}

\usepackage{tikz}
\usetikzlibrary{positioning}
\usetikzlibrary{matrix}

\usepackage{pgf}

\graphicspath{{./images/}{../experiments/images/}}

\renewcommand{\v}[1]{\boldsymbol{\mathbf{#1}}}  % Redefine \v with your desired functionality

\newcommand{\bignorm}[1]{\left\lVert#1\right\rVert}

\newcommand{\Real}{\mathbb{R}}

\newcommand{\ms}[1]{\SI{#1}{\milli\second}}

\newcommand{\framework}{SymX\xspace}

\begin{document}

%\showthe\textwidth  % 510.295pt.
%\showthe\columnwidth  % 243.14749pt

\title{\framework: Energy-based Simulation from Symbolic Expressions}

\author{José Antonio Fernández-Fernández}
\affiliation{%
  \institution{RWTH Aachen University}
  %\streetaddress{1 Th{\o}rv{\"a}ld Circle}
  \city{Aachen}
  \country{Germany}
}
\email{fernandez@cs.rwth-aachen.de}

\author{Fabian Löschner}
\affiliation{%
  \institution{RWTH Aachen University}
  \city{Aachen}
  \country{Germany}
}
\email{loeschner@cs.rwth-aachen.de}
\orcid{0000-0001-6818-2953}

\author{Lukas Westhofen}
\affiliation{%
  \institution{RWTH Aachen University}
  \city{Aachen}
  \country{Germany}
}
\email{l.westhofen@cs.rwth-aachen.de}
\orcid{0000-0003-4427-2377}

\author{Andreas Longva}
\affiliation{%
  \institution{RWTH Aachen University}
  \city{Aachen}
  \country{Germany}
}
\email{longva@cs.rwth-aachen.de}
\orcid{0000-0002-6665-8302}

\author{Jan Bender}
\affiliation{%
  \institution{RWTH Aachen University}
  \city{Aachen}
  \country{Germany}
}
\email{bender@cs.rwth-aachen.de}
\orcid{0000-0002-1908-4027}

\renewcommand{\shortauthors}{Fernández-Fernández, et al.}
 
\begin{abstract}

Optimization time integrators are effective at solving complex multi-physics problems including deformable solids with non-linear material models, contact with friction, strain limiting, etc.
For challenging problems, Newton-type optimizers are often used, which necessitates first- and second-order derivatives of the global non-linear objective function.
Manually differentiating, implementing, testing, optimizing, and maintaining the resulting code is extremely time-consuming, error-prone, and precludes quick changes to the model, even when using tools that assist with parts of such pipeline.

We present SymX\footnote{\url{https://github.com/InteractiveComputerGraphics/symx}}, an open source framework that computes the required derivatives of the different energy contributions by symbolic differentiation, generates optimized code, compiles it on-the-fly, and performs the global assembly.
The user only has to provide the symbolic expression of each energy for a single representative element in its corresponding discretization and our system will determine the assembled derivatives for the whole simulation.
We demonstrate the versatility of SymX in complex simulations featuring different non-linear materials, high-order finite elements, rigid body systems, adaptive discretizations, frictional contact, and coupling of multiple interacting physical systems.

\ch{
    \framework's derivatives offer performance on par with SymPy, an established off-the-shelf symbolic engine, and produces simulations at least one order of magnitude faster than TinyAD, an alternative state-of-the-art integral solution.
}
\end{abstract}




%%
%% Teaser
\begin{teaserfigure}
  \centering
\includegraphics[width=\columnwidth,trim={600 0 600 120},clip]{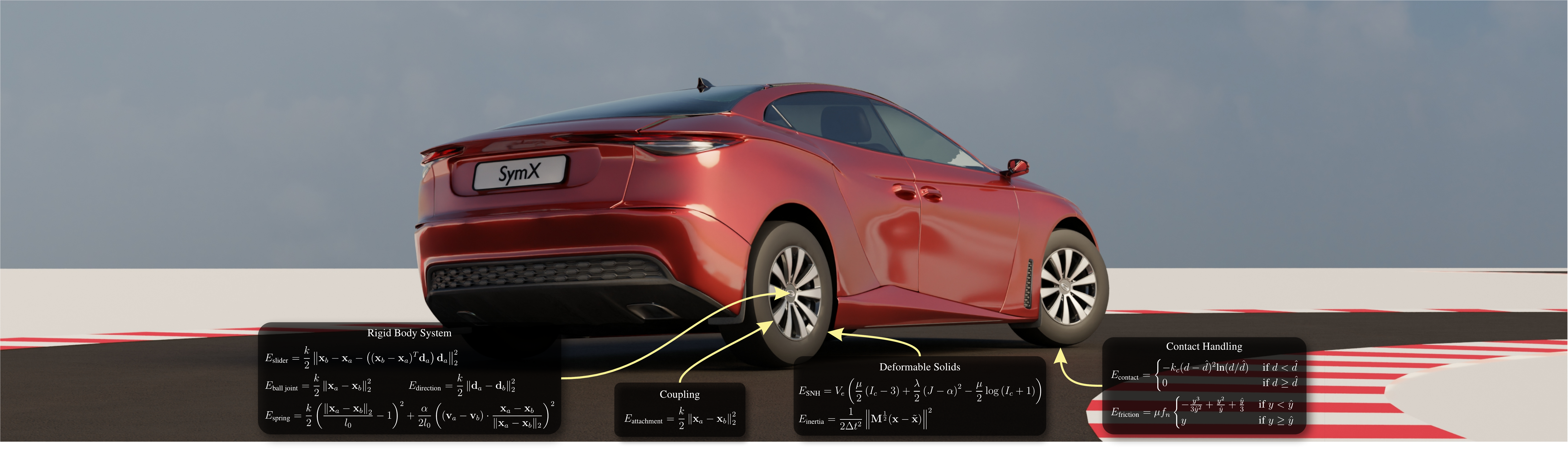}
\caption{Simulation based on an optimization time integrator of a car drifting through a tight hairpin corner with strong coupling between rigid bodies and deformable solids.
  The simulation model consists of nine non-linear potential energies: FEM with linear tetrahedra and the Stable Neo-Hookean material model~\cite{StableNeoHookean} for the tires, constraint-based energies~\cite{MEM+20} for the rigid body components (sliders, ball joints, direction joints, and damped springs), attachment constraints for the coupling of the rigid body system for the suspension with the tires and a frictional contact potential based on the Incremental Potential Contact method~\cite{IPC}.
  All energies are succinctly defined using \framework, which can automatically compute the global gradient and Hessian used to solve the optimization time integration.
}
\label{fig:teaser}
\end{teaserfigure}

\maketitle

%%
%% Sections
\section{Introduction}
\label{sec:introduction}

In the research area of physically-based simulation a common problem is to efficiently compute the solution of non-linear equations, e.g., to simulate non-linear materials~\cite{StableNeoHookean}, to handle collisions with friction~\cite{AEF22}, or to resolve non-linear constraints~\cite{BET14}.
This problem is also highly relevant for simulation methods based on energy minimization which have become increasingly popular in recent years~\cite{GSS+15,NOB16,BOFN18,CLL22}.
Such methods allow the user to combine different material models and constraints in a single simulation by formulating (typically non-linear) potential energy functions for each component. 
Implicit time integration is often performed by minimizing the sum of the inertia energy and all energy potentials, e.g., using Newton's method.
While first-order methods, such as Projective Dynamics~\cite{BML14}, may be used to solve this type of problem, in this work we focus exclusively on second-order methods due to their strong convergence and robustness guarantees~\cite{IPC}.

In this context, first- and second-order derivatives of many, and possibly very complex, energy expressions are required for the minimization process.
Simulations with multiple interacting physical systems, such as rigid and deformable bodies, might require tens of different energies when considering, not only the internal mechanical effects, boundary conditions and joints, but also contacts and friction between their discretization primitives, e.g. triangles, edges and vertices.
Developing, testing and maintaining efficient simulation code to evaluate these energy expressions and their derivatives, and to assemble the results into global data structures is a laborious and error-prone endeavour.

There already exist tools which try to solve some of these problems, e.g., by computing the required derivatives using automatic differentiation or frameworks and languages that assist with the assembly process.
However, there is no tool capable of automating the whole pipeline in an effective manner.
In this work, we propose an integrated solution to differentiation and assembly in the context of physically-based simulation with the following goals:
\begin{itemize}[leftmargin=*]
		\item \textbf{Automation:} First- and second-order derivatives should be computed and assembled completely automatically.
		\item \textbf{Performance:} The evaluation of the energy expression and its derivatives must be fast in order to make the system relevant beyond very early prototyping or small simulations.
		\item \textbf{Productivity:} It should be easy to add or to change expressions and recompilation times should be short to avoid user idling.
		\item \textbf{Flexibility:} 
		The system should work with user-defined data structures, notably the performance-critical sparse matrix for the global Hessian, while imposing no limitations on the choice of minimization method or linear system solver.
		Additionally, it should support dynamic problem topologies and enable the processing of individual element contributions, for example to allow for projection to positive semi-definiteness.
		\item \textbf{Accessibility:} The system should be accessible and uncomplicated to set up, build and distribute in order to further facilitate the exchange of ideas between researchers and the replicability of other's work.
\end{itemize}

In this paper we show in a detailed analysis that existing tools fail to fulfill at least one of these requirements and present our open source framework, \framework, which addresses these points.
By drastically reducing the time spent on differentiation and tedious or repetitive implementation tasks, our proposed system enables researchers to explore ideas very efficiently and to easily compare between different concepts, virtually eliminating iteration delays.
While our system can be used as a prototyping tool, it provides enough performance as-is to run relatively large scenes with complex state-of-the-art models as we show in Fig.~\ref{fig:teaser}.
\framework has already been used in research on complex materials and interactions \cite{micropolar, micropolar_shells, magnetic_rigid}, in differentiable simulation \cite{IFA}, and as the core of STARK \cite{stark}, a simulator for strong coupling between deformable and rigid bodies for use in robotics.
Section~\ref{sec:applications} of this document includes example applications using \framework for non-linear materials, high-order finite elements, rigid body systems, adaptive discretizations, frictional contact, and coupling of multiple interacting physical systems.

\section{Related Work}
\label{sec:related_work}

In this section we first cover simulation methods that require first- and second-order derivatives to guarantee robustness.
Then, we give an overview of the broad landscape of automated approaches to compute derivatives and other systems that make possible to express complex problems in terms of succinct expressions or programs.
We refer to Section~\ref{sec:discussion} for an in-depth discussion about the feasibility of applying specific methods and tools listed herein to our application.

\subsection{Optimization Time Integrators}

Using an incremental potential formulation~\cite{OS99} for dynamic problems is a common approach in computational mechanics.
Derived or related methods also have become popular in computer animation where they are often referred to as optimization time integrators.

Formulating the dynamic systems as a scalar optimization problem instead of a non-linear system of equations was shown to be favorable for robustness and efficiency of the implementation~\cite{KYT06,GSS+15}.
While this robustness is usually associated with Newton-style methods that use a full Hessian, local approaches such as Projective Dynamics~\cite{BML14,NOB16} can be used in case of stricter performance constraints.
To fulfill high accuracy requirements, Li et al.~\shortcite{LGL19} proposed a different method using domain decomposition that improves efficiency especially in case of extreme non-linear and high-speed deformations.
Recently, optimization-based contact models gained considerable popularity.
The Incremental Potential Contact (IPC) approach~\cite{IPC} and its extension to Codimensional IPC~\cite{LKJ21} excel at providing robust interpenetration-free frictional contact handling.
The characteristic robustness and convergence of such methods is subject to having access to second-order derivative information of the underlying global objective function.
While contact potentials with barriers in general appear to be a promising choice for many applications, introducing them to orthogonal phenomenological research projects or existing multi-physics systems~\cite{multiphysics} can require significant development effort.
As we show later, our framework allows users to easily integrate models inspired by IPC in already complex simulation settings.

%While most of the previous works used an incremental potential formulation of backward Euler, Brown et al.~\shortcite{BOFN18} presented a corresponding formulation of the TR-BDF2 integrator as part of a method to more accurately model dissipative forces.

%Beside research to improve optimization-based methods in general, there is also widespread use of such methods for specific applications.
%This includes amongst others example-based elastic material simulation~\cite{MTGG11}, where potentials guide deformations towards example data and crowd simulation which uses potentials to avoid collisions of agents~\cite{KSNG17}.

%In some of the aforementioned publications it is suggested to use Quasi-Newton methods that do not construct a global Hessian for the final implementation.
%Still, they arguably present good examples where our framework could help future researchers in similar positions to evaluate and validate optimization-based approaches without the challenges associated with first-order or Quasi-Newton methods and focus on the modeling aspect itself.

\subsection{Differentiation}
Automating the task of differentiation via computer programs has a long history, the dissertation of John F. Nolan~\shortcite{Nolan1953ADo} being one of the original works in the field.
Over the decades that followed, the relevancy of this field has seen huge leaps forward, and it is at the core of today's most advanced technologies in important fields such as artificial intelligence.
Since it is out of the scope of our work to give an extensive review of the field, we point the interested reader to the book by Griewank et al.~\shortcite{diff_book}.
%Perhaps more closely related to the field of computer graphics is the recent review of Schroeder~\cite{craig_notes}.

There are different strategies to differentiation.
\textit{Automatic differentiation} (AD) is perhaps the most widely used one due to its capabilities to handle derivatives of complex computer programs with dynamic control flow.
In AD, a computation graph of the program to differentiate is built and derivative information is propagated along with the original computation.
At a very high level, AD techniques can be divided in two main categories, \emph{backward} and \emph{forward} mode.
The former one is more efficient when the program has a large number of degrees of freedom, while the latter can be more efficient otherwise.
In recent years, the increased interest in machine learning has brought a lot of attention to backward AD techniques and very powerful tools, such as TensorFlow~\cite{tensorflow} or PyTorch~\cite{PyTorch}, have been widely adopted.
In our setting, however, due the structure of the problem, we need to compute derivatives of local functions which depend on a relatively low number of degrees of freedom, therefore forward mode is usually preferred.
We refer the reader to the work by Schmidt et al.~\shortcite{TinyAD} which presents an in-depth discussion on the efficiency of forward and backward AD for such problems.
The authors also provide an implementation, TinyAD, that is shown to outperform state-of-the-art tools in their applications.
Aside of AD solutions to specific problems, there are general purpose AD tools that can be used to conveniently obtain derivatives in a more general context, e.g. CasADi~\cite{casadi}, albeit at the cost of being unable to fully exploit the structure of the problem at hand.
%This generality however often plays against them when compared with problem-specific solutions as they cannot leverage crucial characteristics of the problem or provide the required problem-specific features.
\chf{
    We also point to Enzyme~\cite{Enzyme}, an LLVM-based AD compiler plugin, and Tapenade~\cite{Tapenade}, a source-to-source AD tool, as further examples of software that can generate efficient derivatives from code.
}

On the other hand, \textit{symbolic differentiation} can be used to generate derivatives from input mathematical expressions.
Dynamic loops and branching are usually more restricted in comparison to AD solutions, but the upside is that there is potentially more room for static analysis and optimization of the expressions, assuming that the target function can be described in closed form.
Symbolic differentiation used as an external tool to the main application (e.g., using SymPy~\cite{sympy}, Mathematica~\cite{Mathematica} or Maple~\cite{maple}) is a well-known option.
This approach has seen some criticism~\cite{craig_notes} related to performance and the error-prone, often manual, process to integrate the generated code into the simulator.
However, efficiency concerns can be addressed by using Common Sub-expression Elimination (CSE) on the resulting derivative expressions, which can be carried out directly in the aforementioned tools.
Recently, the work by Herholz et al.~\shortcite{SymbolicLib} has proven that integrating symbolic differentiation in the application code, coupled with CSE and on-demand compilation can solve the performance shortcomings while making the process completely autonomous.
\ch{Concurrently with our work, Herholz et al.~\shortcite{HSK24} expands on the method by incorporating assembly instead of generating code for the entire problem.}

%Further discussion on the relation between current differentiation approaches and our work is presented in Section~\ref{sec:discussion}.

\subsection{Simulation Systems and DSLs}

In the context of simulation, Domain Specific Languages (DSLs) aim to simplify description and solution of specific problem classes or systems that process or encompass an entire program.
Liszt~\cite{Liszt} is a DSL designed to develop mesh-based PDE solvers that allows to define data at discretization nodes, batching subsequent operations for efficient processing.
Simit~\cite{simit} and Ebb~\cite{Ebb} are DSLs designed to ease writing high performance simulations by splitting the problem definition between data structures and simulation code and automatically generating routines taking care of sparse matrix assembly \ch{both on the CPU and on the GPU}.
More recently, Taichi~\cite{Taichi} and MeshTaichi~\cite{MeshTaichi} take this further by allowing internal data structures to be changed, allowing the user to easily find which is the most suitable for their application.
%Our method is an application specific approach, and as so there are other systems that are more closely related.
DeVito et al.~\shortcite{Opt} proposed a DSL to solve non-linear least squares problems with first-order methods from a concise objective function definition using symbolic differentiation at intermediate representation level.
Further, Thallo~\cite{Thallo} presents performance improvements by allowing computation and storage reorganization of the code.

Outside of DSLs, SANM~\cite{SANM} is a solver that applies the Asymptotic Numerical Method fully automatically to problems defined symbolically.
ACORNS~\cite{ACORNS} generates first- and second-order derivatives of target functions defined in the main application codebase at build time.
Herholz et al.~\shortcite{SymbolicLib} propose a code generator to transform symbolically defined sparse operations into compiled high performance applications that avoid expensive sparse data structure bottlenecks.
Similarly, Dr.Jit~\cite{drjit} compiles per-scene kernels to accelerate execution times in the context of physically-based differentiable rendering.
%All three approaches employ either automatic or symbolic differentiation to internally generate derivatives from the user problem definition.
\framework follows the general philosophy of splitting core definitions from the internal procedures and data structures.
To the best of our knowledge, none of the above methods fulfil all the requirements established in Section~\ref{sec:introduction}: some are too specialized for other applications or require a particular type of solver, and others are not flexible enough in terms of discretization and sparsity or are generally not efficient enough, especially when considering second-order derivatives.

Another class of automated systems are PDE solvers which facilitate the process of using the Finite Element Method (FEM) to solve problems defined in the continuum.
Popular examples are FEniCS~\cite{fenics}, Firedrake~\cite{firedrake}, Freefem \cite{freefem} or Moose~\cite{moose}.
These type of frameworks usually offer a wide range of capabilities such as the use of different finite element spaces, meshing, choice of solver, distributed computing and more.
However, most problems found in computer graphics are (at least partially) discrete in nature (e.g., rigid body dynamics, contacts or friction), rendering this class of frameworks unfit for our task.
%In comparison, \framework stands as a more focused and lower level solution, specialized in formulating and handling the type of problems commonly encountered in computer graphics applications, which are often (at least partially) discrete in nature.

\section{Problem Definition}
\label{sec:method_problem_definition}
Many physical models used in simulation satisfy the following ordinary differential equation
\begin{align}
\label{eq:equations_of_motion}
\v M \dot{\v v} = \v f (\v x) = - \nabla E(\v x), \qquad \qquad \dot{\v x} = \v v.
\end{align}
Here $\v x$ is a vector containing some variant of positional degrees of freedom of the discrete system, $\v v$ similarly contains the velocity degrees of freedom, $\v M$ is the \emph{mass matrix}, \ch{which might be constant or depend on $\v x$}, $\v f$ is a discrete representation of the forces acting on the system and $E$ is a scalar potential function. 
This ODE does not readily hold for rigid bodies without the introduction of a kinematic map~\cite{BET14}, but in the interest of a simpler presentation we leave this aspect out of the present discussion.
In general, dissipative forces, friction for instance, might not have an associated scalar potential $E$ in the formulation above.
In such cases, it is often possible to work around this restriction by \emph{lagging} the dissipative forces in question in some fashion~\cite{IPC}.
% "we incorporate damping at the discrete stage once we have the relation between x and v"

To compute one time step of size $\Delta t$ for this problem, \ch{and without loss of generality,} we may for example use the reformulation of Backward Euler as an optimization problem (cf.\ \cite{GSS+15, NOB16, KKB2018}) to obtain the \emph{incremental potential}
\begin{align}
\label{eq:implicit_euler_ip}
E_{\text{BE}}(\v x) := \frac{1}{2 \Delta t^2} \left \| \v M^\frac12 (\v x - \tilde {\v x}) \right \|^2 + E(\v x)
= E_{\text{inertia}}(\v x) + E(\v x),
\end{align}
where $\tilde {\v x} = \v x(t) + \Delta t \v v(t) + (\Delta t)^2 \v M^{-1} \v f_{\text{ext}}$ and $\v f_{\text{ext}}$ is the vector of external forces which are constant during a time step.
The associated update rules are
\begin{equation}
\begin{split}
	\v x(t + \Delta t) &= \min_{\v x} E_{\text{BE}} (\v x) \\
	\v v(t + \Delta t) &= \frac{1}{\Delta t} \left (\v x(t + \Delta t) - \v x(t) \right ).
\end{split}
\label{eq:implicit_euler}
\end{equation}
\chf{Note that many other integration methods permit a similar reformulation as an optimization problem, such as the midpoint rule~\cite{DLK18}, the trapezoidal rule, BDF2 and TR-BDF2~\cite{BOFN18,CLL22}}.
We can describe the associated minimization problem as a sum of energy functions
\begin{align}
\label{eq:problem_definition_terse}
\min_{\v u} \sum_i E_i(\v u; \mathcal{P}_i)\,
\end{align}
in which we have used the state vector $\v u$ to describe the degrees of freedom, typically positions or velocities. % as for some applications it may be favorable to pose the minimization problem either in terms of positional or velocity degrees of freedom.
The abstract quantity $\mathcal{P}_i$ represents the parameters of the energy function $E_i$, i.e., the data of the problem that is not dependent on the state $\v u$.

Usually, energies can be decomposed into a number of smaller contributions.
For example, the total strain energy $E_{\text{strain}} = \sum_e E_{\text{strain}, e}$ for a deformable finite element model is the sum of the individual element strain energies $E_{\text{strain}, e}$.
To capture this inherent structure of the problem, we introduce abstract \emph{elements} to the formulation.
In practice, an element is an entity that has a contribution to the global potential energy, e.g., a tetrahedral finite element to simulate a deformable solid, a rigid body or a contact point between two objects.
Each energy $E_i$ then gets associated with a set of elements $\mathcal{E}_i$ where it is defined and evaluated.
We now replace Eq.~\eqref{eq:problem_definition_terse} with our general problem formulation
\begin{equation}
    \label{eq:problem_definition}
    \min_{\v u} E(\v u) = \sum_{i} \sum_{e \in \mathcal{E}_i} E_i(\v R_e \v u; \mathcal{P}_{i,e})\,,
\end{equation}
where $\v u$ denotes the \emph{global} degrees of freedom and $\v R_e$ is the selection operator that extracts the degrees of freedom specific to the element~$e$.
$\mathcal{P}_{i,e}$ are the parameters specific to element~$e$ for energy $i$.
In other words, $\v R_e$ maps global to \emph{element-local} quantities, and in consequence an energy $E_i$ operates only on element-local inputs of the same size. Its definition is independent of a specific element \chf{instance}; only the parameters change. 
\chf{This concept can also be applied to models that require (possibly non-linear) mappings, e.g. between coordinate systems for rigid bodies, by simply folding such mappings into the definition of $E_i$ and defining different sets of elements $\mathcal E_i$ with an associated energy for different compositions of mappings.}
%\jf{I think this is what incites confusion about linear mappings.}

To summarize, each energy function $E_i(\hat{\v u}; \mathcal{P}_i)$ is therefore a function of a generic vector $\hat{\v u}$ with \emph{fixed} input size, evaluated for each associated element in $\mathcal{E}_i$.
It is then possible to symbolically represent, differentiate and generate code for each $E_i$, and finally assemble the overall derivative of $E$ by summation.

% Probably don't need these expressions, but keeping them here as they're annoying to write out again
%The gradient and Hessian of $E$ are given by
%\begin{align}
%\nabla E &= \sum_i \sum_{e \in \mathcal{E}_i} \v R_e^T \nabla_{\hat{u}_i} E_i, \\
%\v H &= \frac{\partial^2 E}{\partial \v u^2} = \sum_i \sum_e \v R_e^T \v H_{i,e} \v R_e.
%\end{align}

Under the assumption that $E(\v u)$ is at least $C^1$ continuous, we can efficiently solve \eqref{eq:problem_definition} with an appropriate choice of optimizer (see~\cite{Nocedal}).
%First-order optimization methods require the gradient $\nabla E$, and second-order methods require the Hessian $\v H$ as well.
%For particularly challenging problems --- such as those involving stiff materials or challenging contact --- first-order methods may converge too slowly, and second-order methods are preferable, such as variants of Newton's method.
%Although our method can be used with first-order methods, it is substantially more difficult to obtain efficient second-order derivatives, and therefore our framework brings even more to the table when used with second-order optimizers.

\subsection{Example: Deformable solids}
\label{sec:example_deformable_solids}
We now demonstrate how to formulate a motivating example within the mathematical framework of \eqref{eq:problem_definition}.
We wish to simulate a deformable solid with the non-linear Neo-Hookean material using a linear tetrahedral finite element discretization and the Backward Euler integrator, subject to gravity.
We let $\v u = \v x$ be the global vector of deformed vertex positions, and each element is associated with four vertices, forming a local vector $\hat{\v x} = \v R_e \v u = \v R_e \v x \in \mathbb{R}^{12}$ containing the deformed vertex positions stacked in an element-local vector.
From this we can compute the deformation gradient $\v F_e = \v F_e(\hat{\v x})$ of the element~\cite{SB12a}.%, which is constant across the element for the case of linear elements.

%In general, the strain energy density $\psi = \psi(\v F)$ for a hyperelastic material model depends only on the deformation gradient $\v F$. 
The strain energy density for the Neo-Hookean model is given by
\begin{equation}
    \label{eq:neohookean}
    \psi^\text{NH} = \frac{\mu}{2}(I_c - 3) + \mu \text{log}(\text{det}(\v F)) + \frac{\lambda}{2}\text{log}^2\left(\text{det}(\v F)\right),
\end{equation}
where $\mu$ and $\lambda$ are the Lamé parameters and $I_c = \text{tr}(\v F^T \v F)$~\cite{StableNeoHookean}.
%With the deformation gradient and strain energy density in hand, 
We can compute the strain energy for the element by integrating the strain energy density over its domain $K_e$
\begin{align}
E_{\text{NH},e} (\hat{\v x}) = \int_{K_e}  \psi^{NH}(\v F) \mathrm{d} \v X
= V_e \, \psi^{NH}(\v F_e(\hat{\v x})).
\end{align}
Here $V_e$ denotes the volume of the element.
%We can formulate the inertia energy necessary for the Backward Euler incremental potential \eqref{eq:implicit_euler_ip} in a similar fashion, and our total energy function for the minimization problem \eqref{eq:problem_definition} becomes
Our total energy function for the minimization problem \eqref{eq:problem_definition} becomes
\begin{align}
E(\v x) = E_{\text{inertia}} + \sum_e E_{\text{NH},e}(\v R_e \v x).
\end{align}
Since the gradient and Hessian are computed and assembled by our framework, only the energy functions $E_{\text{inertia}, e}$ and $E_{\text{NH},e}$ need to be provided in symbolic form by the user.
\section{Existing solutions}
\label{sec:discussion}

% \begin{figure}[tb]
%     \centering
%     \includegraphics[width=\columnwidth,trim={0 0 0 0},clip]{images/enhanced_nh.png}
%     \caption{\todo{A very soft but hardly stretchy viscoelastic sphere is dropped on a plane and is let some time to slowly deform into a pancake-like shape. Then, a prescribed rigid box descends, presses the viscoelastic material, and is twisted. Finally, the box ascends, revealing a footprint left on the viscoelastic deformable object.}}
%     \label{fig:enhanced_nh}
% \end{figure}

\ch{
    Our framework, is designed to facilitate the exploration of novel, complicated potentials and intricate interactions across multiple systems and discretizations.
    While \framework can of course implement relatively simple simulations and well-known potentials found in the literature, its principal advantage lies in supporting work beyond that.
    %Two of our goal applications, and where \framework excels, are (a) when the number of distinct potentials is large, as it is the case with multi-system interactions, e.g. the car scene in Fig.~\ref{fig:teaser} or the dryer simulation in Fig.~\ref{fig:dryer}, and (b) when material formulations become very involved, such as for the Enhanced Neo-Hookean potential shown in Fig.~\ref{fig:enhanced_nh} which features strain-rate viscoelasticity and and volumetric strain limiting.
}

\ch{
    With that objective in mind, we now examine existing approaches to differentiation, evaluation, and assembly in the context of our problem as defined in Eq.~\eqref{eq:problem_definition}, and compare their suitability against the requirements of automation, performance, productivity, flexibility, and accessibility set forth in Section~\ref{sec:introduction}.
}

\subsection{Manual implementation}
\label{sec:manual}
The baseline option is to differentiate the energies by hand and to manually implement and optimize the corresponding evaluation and assembly.
\ch{Naturally, in the context of commonly used potentials this can be relatively straightforward since the derivatives might be known, but this is not always the case in research.}
Thorough manual code optimization can yield very high performance results and the approach is flexible, however, manual implementations are typically very time-consuming and error-prone to develop, test and maintain.
Moreover, changes or additions to the existing energies are slow, impeding fast prototyping of new solutions.

\subsection{Numerical differentiation}
While numerical differentiation has seen impressive advances in robustness and can provide reliable derivative information, for example with the Complex Step method~\cite{complex_step}, it still faces the fundamental problem that it requires multiple evaluations of the energy value itself, at least one for each entry in the gradient and Hessian.
In our testing on a linear tetrahedral element with a $12\times12$ Hessian, evaluating the value of the energy was significantly more expensive than 1/144th of the runtime needed for the whole Hessian matrix.
Therefore, we consider numerical differentiation unsuitable for our application.

\subsection{Automatic differentiation}
AD is often the solution of choice for many applications due to its flexibility, easiness of integration in existing codebases and large capabilities for automation, which is why it is commonly used for prototyping and testing.
AD excels at differentiating complex programs with arbitrary control flow that depend on a large number of variables.
Our problem, however, has a very specific structure (Eq.~\eqref{eq:problem_definition}) which can be leveraged for efficient generation and evaluation of derivatives.
General differentiation frameworks, e.g. CasADi~\cite{casadi}, cannot take advantage of the structure of our specific problem, necessitating a formulation of the global energy as an explicit sum of all element contributions to differentiate with respect to all global (instead of local) degrees of freedom, which becomes unfeasible at large scales.
Additionally, the limitations of general-purpose AD tools are further exacerbated when the topology of the problem changes, for example due to dynamic contacts or remeshing, necessitating the recalculation of global derivatives and/or problem sparsity.
%The advantage of \framework in this regard is that it only differentiates unique energy functions (in the order of tens), and evaluates the results per element contribution, as oppose to differentiating all the element contributions (thousands or millions).
The lack of structural awareness of the problem also inhibits the possibility to perform per-element operations, such as projecting element Hessians to the cone of positive semi-definite matrices, which is a common practice in second-order minimization frameworks~\cite{TSIF05, IPC}.

After evaluating various tools, we have determined that TinyAD \cite{TinyAD} is the best AD candidate for our problem as it is specifically designed to compute the same type of derivatives found in our applications, supports per-element projections to positive semi-definiteness, and, as the authors show in their original paper, it outperforms established AD libraries for such problems.
To assess if TinyAD meets our performance requirements, we conduct comprehensive performance comparisons in Section~\ref{sec:results}.

\subsection{Symbolic off-the-shelf tools}

Symbolic mathematical engines such as Mathematica~\cite{Mathematica}, Maple~\cite{maple} or SymPy~\cite{sympy} can be used to compute derivatives and generate corresponding code.
However, our proposed framework not only performs the differentiation and code generation, but it is also aware of the simulation data structures and as such is able to take care of the evaluation of these functions as well as the assembly of the global gradient and Hessian.

\begin{figure}[tb]
    \centering
    \includegraphics[width=\columnwidth,trim={0 0 0 0},clip]{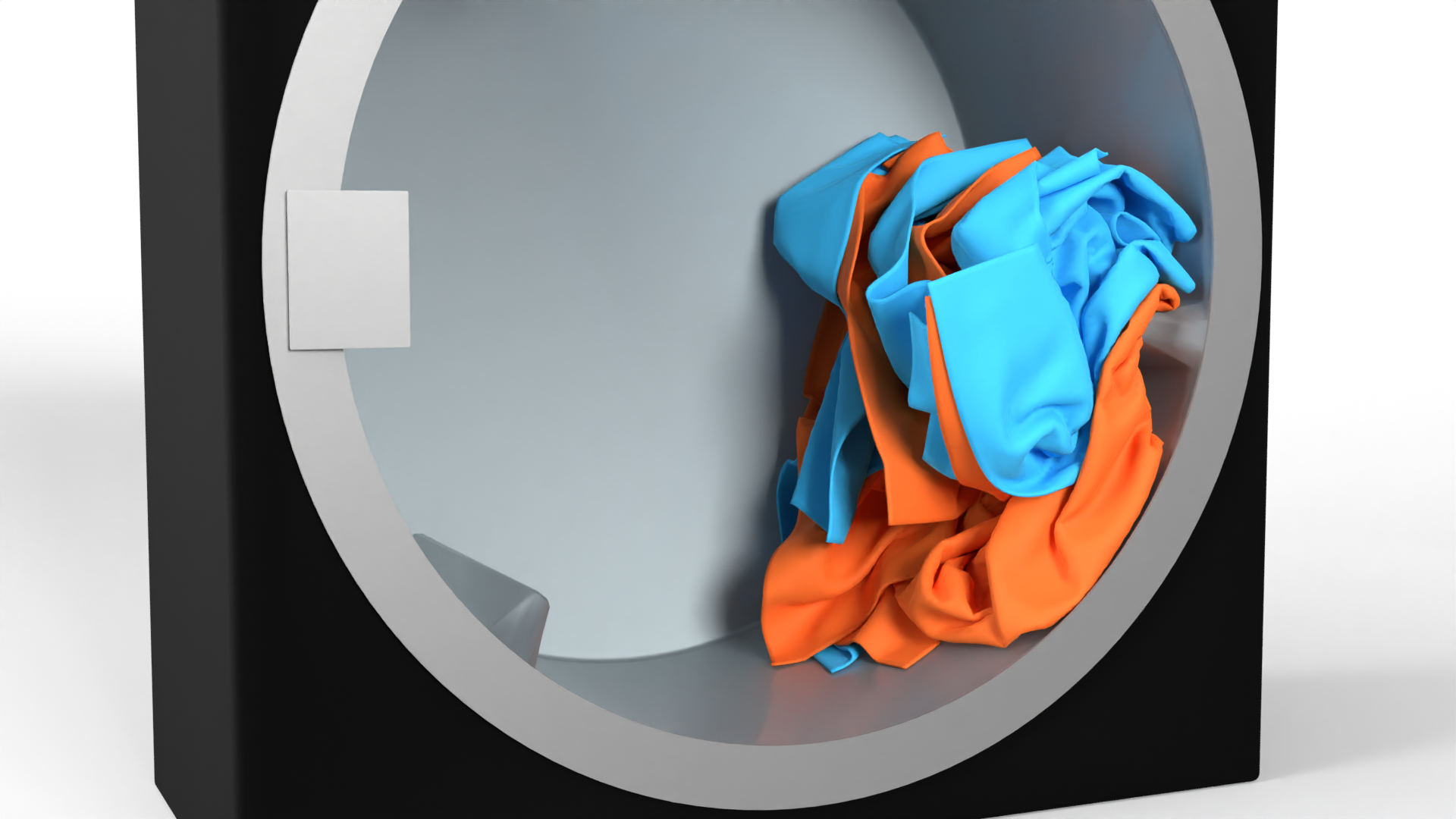}
    \caption{
		The drum of a tumble dryer rotates with eight pieces of cloth inside.
		This scene features a total of 46 distinct energies \ch{(138 auto generated functions), including rigid body dynamics and constraints, shell mechanics and contact and friction potentials for all the combinations between all discretization primitive pairs. The simulation features 245k degrees of freedom.}}
    \label{fig:dryer}
\end{figure}

To highlight why this is desirable, consider the tumble dryer simulation shown in Fig.~\ref{fig:dryer}.
This simulation requires 46 distinct energy types to model deformable materials, rigid bodies, joints and constraints, as well as contact and friction between all discretization primitives.
\framework not only generates and compiles the three required functions per expression to compute the energy, gradient and Hessian (a total of 138 functions), but it is also able to autonomously evaluate them using user-defined accessible data arrays to assemble the global data structures.

Relying solely on external differentiation tools would require the user to generate the code for all the involved per-element energies, followed by manual integration into the simulation code.
To incorporate the externally generated code, the user then has to write glue code for gathering the locally required values for each energy from the global data arrays.
\ch{In the dryer example for instance, this becomes very tedious since each one of the 138 functions has a unique signature and operates on a distinct set of inputs and outputs which requires a function-specific mapping from the simulation data and assembly to the global derivative data structures.}
Even after this initial setup, changes to the expressions might happen regularly in research projects which would require re-running the external tools and potentially updating the function handling in the simulation codebase.
This process is error-prone and time-consuming and therefore does not meet the goals established in Section~\ref{sec:introduction} for automation and productivity.

Some existing symbolic engines, such as Mathematica, provide low-level C interfaces to access their symbolic functionalities, which could be used to avoid relying on external scripts and to integrate the energy definitions directly in the simulation codebase.
However, introducing external differentiation tools in the pipeline still requires implementing the declaration, evaluation and assembly components.
Another problem is that some general purpose tools are not built with performance as a priority, e.g. SymPy is written in Python, \ch{and relying on them for the derivatives can drastically slow down the entire pipeline.
See Section~\ref{sec:results} for SymPy differentiation benchmarks}.
Finally, coupling commercial engines (e.g., Maple or Mathematica) directly into the simulation codebase invalidates our goal of accessibility as defined in Section~\ref{sec:introduction} since closed source licensed software prevents researchers from exchanging ideas or reproducing other works freely. 
In contrast, simulation-native open source solutions such as TinyAD or \framework offer a much more lightweight, fully-automated, single-codebase pipeline and have almost no setup and distribution barriers as only a C++ compiler is required.

\subsection{Simulation systems and DSLs}
While there is a plethora of relevant systems and Domain Specific Languages (DSL) as outlined in Section~\ref{sec:related_work}, we did not find a solution that fulfills all of our requirements.

The most relevant approaches in the context of computer graphics that support second-order derivatives are ACORNS~\cite{ACORNS} and the method by Herholz et al.~\shortcite{SymbolicLib}.
ACORNS can generate Hessians that must be then manually integrated in the simulation but it does not support dynamic branching and is outperformed by integrated solutions such as TinyAD~\cite{TinyAD}.
The method by Herholz et al., on the other hand, presents in fact very good performance by reducing and compiling all the sparse queries into a single program, but this prohibits changes in the sparsity pattern, necessary for contacts and remeshing.
Further, their method has very long code generation and compilation times as they show in Table 3 of their work~\cite{SymbolicLib}.

\ch{
    Considering solutions with no support for differentiation, we find that Simit~\cite{simit} and Ebb~\cite{Ebb} are effective simulation systems that offer great convenience and performance.
    Assuming that the derivatives of all the potentials needed in the simulation are known, these systems offer scripting languages that allow the user to conveniently describe such potentials, as well as other components of the simulation, such as time-stepping schemes, minimizers and linear solvers.
    While the lack of differentiation capabilities makes them incompatible with our goals, we validate the performance of \framework in a comparison with Simit for the evaluation and assembly of the Neo-Hookean potential energy in Section~\ref{sec:simit_comparison}.
}

\subsection{Conclusion}
We observe that prevailing trends in frameworks for computer graphics \cite{TinyAD} and related fields such as rendering \cite{drjit,Thallo}, machine learning \cite{tensorflow,PyTorch} or mathematics \cite{fenics,firedrake} show that modern tools have proliferated precisely because they are accessible, offer a very high degree of automation and safety and provide appealing flexibility and performance.
The ongoing scientific research into better, problem-specific solutions for complex differentiation applications demonstrates that differentiation is in practice still an open problem, and that existing general purpose tools, while useful, do not offer a definitive solution for all differentiation needs.
In this context, we identify a space for a framework to support researchers in developing and sharing simulation models in the context of Newton-type solvers, which is the motivation behind \framework.

\section{\framework Framework}
\label{sec:method}

\framework is an integrated solution that seeks to fulfil all the requirements defined in Section~\ref{sec:introduction} while avoiding the shortcomings of the existing methodologies.
Note that the system's concern is to provide global assembled derivatives, therefore it does not impose any requirements to the simulation software and it is independent of the rest of its components, e.g. minimization method, \ch{time discretization}, collision detection, etc.
We give a general overview of the framework in Section~\ref{sec:overview}.
The symbolic engine is introduced in Section~\ref{sec:symbolic_engine}.
Finally, we discuss the required matrix assembly in Section~\ref{sec:assembly}.
%While we use a specific application example in this section to explain the components of our framework in detail, it is not limited to this use case and more complex applications are discussed in Section~\ref{sec:applications}.

\subsection{Overview}
\label{sec:overview}

\begin{figure*}[tb]
    \centering
    \includegraphics[width=\textwidth]{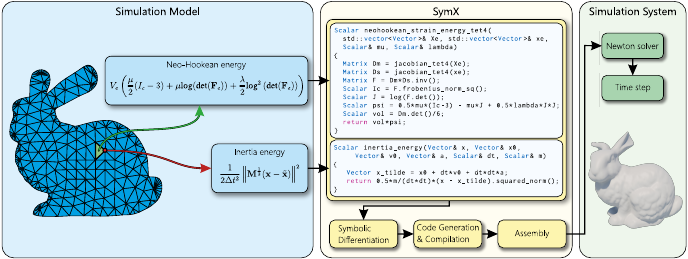}
    \caption{Overview of a simulation step with the \framework framework. Left: The input is a discretized model and energy functions. In our application example we use a tet mesh and the inertia and Neo-Hookean strain energy functions. Center: The user has to implement a symbolic definition of these functions. The framework will then compute the element gradients and Hessians by symbolic differentiation, generate and compile efficient code, and assemble the element contributions to get the global gradient and Hessians. Finally, these terms can be used in a Newton solver to perform a simulation step for the deformable bunny. }
    \label{fig:overview}
\end{figure*}

At a high level, the input to \framework is a collection of symbolic expressions with symbols associated with user-owned data arrays, and it returns the global gradient vector and the global Hessian sparse matrix for the specified sets of degrees of freedom.
Fig.~\ref{fig:overview} shows how the minimization problem of the deformable solids example in Section~\ref{sec:example_deformable_solids} is solved using our framework.
For the simulation model on the left, the user has to implement the mathematical expressions (center). 
Then our framework generates an expression graph and determines the derivatives using symbolic differentiation.
For efficient evaluation, \framework generates source code for each energy, compiles and caches it before the simulation starts.
During the simulation, at the user's request, the compiled functions are evaluated for each energy and for each element and the global gradient and Hessian are assembled.
Evaluation and assembly are automatically parallelized across elements.
See Fig.~\ref{listing:global_energy} for a self-contained \framework example of the setup required to declare the energies defined in Fig.~\ref{fig:overview}.
%Finally, the user can combine several expressions that should be minimized and link their simulation data and connectivity information with the symbols used in the expressions (see Fig.~\ref{listing:global_energy}).
%This enables our framework to autonomously loop over all elements, gather the data, evaluate the compiled expressions, and assemble the results.

%In the following, we explain the individual steps for our application example in more detail.

\begin{figure}[tb]
    \begin{mycodescriptsize}{}
   // Simulation data (uninitialized for brevity)
   std::vector<std::array<double, 3>> x, x0, v0, a, x_rest;
   std::vector<double> lumped_mass;
   double time_step, mu, lambda;
   std::vector<std::array<int, 4>> tets;
   std::vector<int> nodes;

   // Create global energy and define contributions
   GlobalEnergy G;
   DoF dof = G.add_dof_array(x);
   G.add_energy("neo_hookean_tet4", tets,
      [&](Energy& E, Element& tet)
      {
         // Create local symbols from the data arrays
         std::vector<Vector> xe = E.make_dof_vectors(dof, x, tet);
         std::vector<Vector> Xe = E.make_vectors(x_rest, tet);
         Scalar m = E.make_scalar(mu);
         Scalar l = E.make_scalar(lambda);
         
         // Define energy
         E.set(neohookean_strain_energy_tet4(Xe, xe, m, l));
      });
   G.add_energy("inertia", nodes,
      [&](Energy& E, Element& node)
      {  
         // Create local symbols from the data arrays
         Vector xn = E.make_dof_vector(dof, x, node);
         Vector x0n = E.make_vector(x0, node);
         Vector v0n = E.make_vector(v0, node);
         Vector an = E.make_vector(a, node);
         Scalar mn = E.make_scalar(lumped_mass, node);
         Scalar dt = E.make_scalar(time_step);
         
         // Define energy
         E.set(inertia_energy(xn, x0n, v0n, an, dt, mn));
      });
   
   // Compilation
   G.compile("path/to/codegen/directory");

   // Assemble global data structures
   Assembled assembled = G.evaluate_E_grad_hess();

    \end{mycodescriptsize}
    \caption{\framework code to define, compile and evaluate the inertia and strain energies as well as their gradients and Hessians for the example problem defined in Fig.~\ref{fig:overview}.}
      %Example code to link simulation data to \framework's symbols for the evaluation of the Neo-Hookean strain energy and the inertia energy, together with their respective first- and second-order derivatives, as defined in our problem example.
    \label{listing:global_energy}
\end{figure}

\subsection{Symbolic Engine}
\label{sec:symbolic_engine}

The core of our framework is the symbolic engine.
As input, the engine requires a symbolic mathematical expression for each energy, which is done using \framework's symbolic types \texttt{Scalar}, \texttt{Vector} and \texttt{Matrix}.
The framework provides operator overloading and common linear algebra functionalities for these types.
\ch{
    Both, the symbolic \texttt{Vector} and \texttt{Matrix}, are dynamically allocated arrays of scalar expressions.
}
%Fig.~\ref{fig:overview} shows the symbolic expressions of the Neo-Hookean potential energy and the inertia energy for the application example described in Section~\ref{sec:example_deformable_solids}.

Instead of directly executing the operation of a symbolic expression, our engine internally generates an expression graph \chf{of scalar expressions}.
In this graph, each node represents either a user-defined symbol, a constant value or an operation applied to the result of its child nodes.
\framework supports arithmetic and trigonometric operations as well as square roots and logarithms, and new operations can be added.
It is also possible to add custom scalar derivative rules.
However, vector and matrix differentiation rules are not yet supported.
Conditional branching is a special type of operation that is discussed in Section~\ref{sec:extensions}.% \todo{move this to limitations and clarify its possible as future work}.

In the following we describe the components of our symbolic engine using the example presented in Section~\ref{sec:example_deformable_solids}.
%However, keep in mind that our framework can also handle far more complex configurations (e.g., coupling multiple materials, using high-order elements, handling collisions etc.) as we will show in Section~\ref{sec:applications}.

\subsubsection{\chf{Common subexpression elimination}}
\chf{
    A naively constructed expression graph of all scalar operations for common energies typically contains many reoccurring identical subgraphs.
For example, consider a single entry of the deformation gradient of a Lagrangian finite element.
The corresponding symbolic expression becomes quite complex for higher-order elements and occurs multiple times in typical strain energy densities.
We identify and eliminate structurally identical sub-graphs in order to reduce the amount of generated code and improve performance of our symbolic differentiation.
To achieve this, we use a hash map local to each energy that stores all expression nodes that were already created in the graph together with an identifier.
Whenever a new expression is added, we perform a lookup and replace it with an existing identifier if an identical expression was previously constructed.
This deduplication is performed in a bottom-up way when expressions are constructed, therefore it is sufficient for the lookup of an expression to only compare the expression type and identifiers of its direct children expressions to guarantee uniqueness.
In our example, the expression complexity of the Neo-Hookean potential Hessian defined on a linear tetrahedral element (see Fig.~\ref{fig:overview}) was reduced by 70\%, from 7517 to 2284 operations.
}

\chf{To also eliminate algebraically equivalent expressions,~\citet{SymbolicLib} proposed ``algebraic hashing'' which assigns hashes to elementary nodes (e.g. variables and literal numbers) and computes hashes of more complex expressions by applying their corresponding operations such as multiplications and additions to the hashes of their subexpressions.
However, while improbable, it is possible that the operations performed on the hashes lead to hash collisions of non-equivalent expressions.
%To resolve this, it would be required to actually prove or disprove the algebraic equivalence of the expressions which is out of scope of these frameworks.
Therefore, we instead opt for the more robust approach of checking for structural identity.}

%\jf{Do we mention that we observe little difference in performance between hashing methodologies? As it is, it sounds like we have decided to not do a lot of simplifications to the math!}

\subsubsection{Symbolic Differentiation}
\chf{
    To compute the derivatives of a symbolic expression, our framework recursively traverses its expression graph and \chn{applies} the chain rule with table lookups for the derivatives of elementary functions.
    During this traversal, we use a different hash map to cache derivatives of subexpressions that were already computed, making use of the previously performed common subexpression elimination.
    The same cache is used across the entries of an element gradient and Hessian as their expression trees often have significant overlaps which significantly reduces the time required for differentiation.
    In the case of the Neo-Hookean energy potential defined on a linear tetrahedra element, differentiation times improve from $\SI{1.88}{\milli\second}$ to $\SI{0.44}{\milli\second}$ thanks to caching intermediate derivatives.
}

% =====================================================================================================================
\subsubsection{Code Generation and Compilation}
Once we have the symbolic expressions for a function and its derivatives, we need to evaluate them for all the elements.
However, evaluating the expressions by traversing the expression graph would be prohibitively slow, instead, equivalent C++ code for such functions is generated and compiled, which can be hundreds of times faster.
\ch{
    To this end, the expression graph is traversed bottom-up, collecting all operations in the order they need  to be calculated, emitting one line of C++ code per operation or graph node.
    Every generated function has two arguments: a pointer to an input buffer, with the data corresponding to the input symbols (e.g. element vertices, material parameters, etc.), and a pointer to an output buffer (e.g. energy value, gradient and Hessian).
    Fig.~\ref{listing:generated_code} partially shows the generated function to evaluate the energy value, gradient and Hessian of the Neo-Hookean energy defined in Fig.~\ref{listing:global_energy}.
    It has 26 inputs (12 for \texttt{xe}, 12 for \texttt{Xe}, 1 for \texttt{mu} and 1 for \texttt{lambda}), and 157 outputs, (1 for the energy value, 12 for the gradient and 144 for the Hessian).
    \framework manages the buffers and the mappings between the symbols and the inputs and outputs of the generated functions.
    Besides the function itself, each generated C++ function file contains some metadata with a signature hash id, and the number of inputs and outputs.
}

\begin{figure}[tb]
\begin{mycodeblock}{}
  void neohookean_tet4_hess(double* in, double* out)
  {
    /*
       Add:     450
       Sub:     310
       Mul:     1347
       Inv:     3
       PowN:    11
       Log:     1
       Total:   2122
    */
    double v49 = in[46] * 0.5;
    double v51 = in[24] * in[45];
    double v52 = in[12] + v51;
    ...
    out[144] = v1898;
    out[145] = v931;
    out[146] = v1088;
  }
\end{mycodeblock}
\caption{\ch{Generated C++ function to evaluate the energy value, gradient and Hessian of the Neo-Hookean energy for a tet4 element from the definition in Fig.~\ref{listing:global_energy} lines 11 to 22.
Only some representative lines from the top and the bottom of the function are shown.}}
\label{listing:generated_code}
\end{figure}

% =====================================================================================================================
\subsubsection{Data-Symbol Mapping}

A central concept in \framework is the mapping between simulation data and its corresponding symbols, as this key design principle allows it to evaluate the generated functions and to assemble the global structures autonomously.
%This key design decision allows the system to be easily integrated into the simulation software and eliminates the need for interfaces between simulation logic, evaluation and assembly.
To achieve this, every symbol in an energy definition is associated with a C++ lambda function that returns an updated view to the data array it represents.
This mapping works for both simulation data (e.g., positions, velocities,~\dots) and connectivity information (e.g., tetrahedra, triangles, edges,~\dots).
Before the evaluation of a specific energy for all related elements, \framework requests updated views of all data arrays associated to it (including its connectivity array) using the lambdas.
The indices represented in each element are used to index the data arrays and to calculate the global indices where the local gradient and Hessian must be assembled at.
The only requirement for an array to be compatible with our system is that it must hold its data contiguously in memory so that it can be accessed by beginning, stride and index.
%Lambda functions to view the arrays are automatically generated for C++ Standard Template Library \texttt{vector} and \texttt{array} and other popular types such as Eigen vectors and matrices~\cite{Eigen}.

We now analyze the code example shown in Fig.~\ref{listing:global_energy}.
% An example of the setup required to declare the energies defined in Fig.~\ref{fig:overview} in our system can be seen in Fig.~\ref{listing:global_energy}.
First the array of degrees of freedom must be declared (line 10) in order to identify the symbols the system will take derivatives with respect to.
Lines 11 and 23 define the energy functions for the tets (strain) and nodes (inertia), respectively.
Within the body of each energy definition, symbols are created as counterparts of the arrays they represent (lines 15-18 and 27-32) and the symbolic expression to evaluate the energy for an element is set (lines 21 and 35).
In line 39, \framework is instructed to compile all the required functions, including derivatives, and to write the shared objects (\texttt{.dll} or \texttt{.so}) in a specific folder.
There are three global evaluation functions available.
From lighter to heavier in regards to runtime: one to compute the global energy (typically used during line search), another to compute the global energy and gradient (to check for Newton's method convergence) and another, used in line 42, to compute the global energy, gradient and Hessian (to assemble the linearized system of equations in Newton's method).
After the initial definitions and compilation, these evaluation functions can be used as many times as necessary.
Every global assembly will be executed using updated user data from the mapped arrays, even when they change in size, for example, in the case of mesh refinement.

\framework also exposes a low-level API to the symbolically generated and compiled functions for cases where manual evaluation of such functions is preferred.
Through this interface, compiled functions can be called directly by the user specifying the data corresponding to each symbol in the expression, bypassing the need for the data-symbol mapping.
It is important to note that while this option grants more control, it reintroduces significant complexities, optimization concerns, and safety responsibilities that \framework is designed to handle automatically.
For the rest of this document it is assumed that the high-level data-symbol mapping is used and that evaluations and assembly are automated.

% =====================================================================================================================
\subsubsection{\ch{Dynamic topology}}
\label{sec:dynamic_topology}

\ch{
    Contact interactions and adaptive mesh refinement are two factors that can lead to changes in a problem's topology.
    For instance, points that were previously apart can briefly come into contact and then separate again.
    Also, new smaller elements may be introduced in regions undergoing large deformations.
}

\ch{
\framework supports dynamic topology in a straightforward, general manner: 
it evaluates and assembles the derivatives for the elements present in the connectivity arrays at the moment of each evaluation request.
This approach works because \framework has updated view access to the data arrays, including their sizes, which are defined and maintained by the user, as explained in the previous section. 
}

\ch{
Hence, the user only needs to update the list of contact pairs or mesh elements (if needed) before calling \framework.
The resulting global derivatives will then reflect the latest state of the simulation, including any changes in the sparsity pattern or even the total number of degrees of freedom.
Further details on dynamic topology assembly can be found in Section~\ref{sec:assembly}.
}

\ch{
It is worth noting that \framework itself does not explicitly define concepts such as collision, contact, or even mesh.
Everything is instead constructed from symbolic expressions associated with element lists, regardless of whether they represent contact pairs, FEM elements, rigid bodies, or something else entirely.
}

% Contacts and adaptive mesh refinement are two of the causes for a problem to change its topology.
% Points separated from each other might get in contact for a brief moment and then separate again and new finer elements can be added in regions of the simulation where the material is subject to large deformations.
% \framework supports dynamic topology (see Section~\ref{sec:applications}) in a very simple and general way: it will evaluate and assemble the derivatives for the elements located in the connectivity arrays --- defined and held by the user --- at the time of the request.
% This is possible because \framework has an updated view to the user arrays, as explained in the previous section.
% Therefore, the user only needs to update the list of contact pairs or mesh elements (if necessary) before calling \framework.
% The resulting global derivatives will represent the updated state of the simulation, including potential changes in the sparsity pattern and even the number of degrees of freedom.
% See Section~\ref{sec:assembly} for details on dynamic topology assembly.
% This is a good place to remark that there is no concept of collision, contact... or even mesh in \framework.
% Everything is constructed from symbolic expressions associated with lists of elements, regardless of whether they represent contact pairs, FEM elements, rigid bodies, etc.

% =====================================================================================================================
\subsubsection{Extensions}
\label{sec:extensions}

In the following we introduce features of the system which are required for more complex simulations or to improve its performance.

% --------------------------------------------------------------------------------
\paragraph{Branching}
\framework supports differentiation and code generation of expressions with arbitrary nested branching using
\begin{mycodeblock}{}
   Scalar res = branch(Scalar& c, Scalar& a, Scalar& b);
\end{mycodeblock}
\noindent
where $c$ is an expression that represents a conditional variable, the expression $a$ is used if $c \geq 0$, and $b$ is used otherwise.
\ch{
    \texttt{branch} emits an actual \texttt{if-else} statement in the C++ generated code and therefore only the correct branch is executed.
    Differentiation does not affect the branching points since the condition stays the same, the only difference is that code to evaluate the derivatives appear on each side of the branch.
}

\ch{
    Branching is extensively used in simulation.
    In this work we make use of it, for example, in the implementation of the mollifiers needed for the IPC edge-edge contacts and the friction potentials.
    Additionally, thanks to branching we can implement expressions with functions like \texttt{min}, \texttt{max}, \texttt{abs} and \texttt{sign}, which allowed us to write the signed distance function to a cylinder, used in the scene shown in Fig.~\ref{fig:armadillo}.
}

% --------------------------------------------------------------------------------
\paragraph{Conditional evaluations}
%\label{sec:conditional_energies}

All the branches spawned by \texttt{branch} will generate results that will be assembled.
However, a special case of branching that needs dedicated treatment is when the value of one branch is zero and therefore derivatives and assembly should be skipped, e.g., when modeling contact barrier potentials.
In this case, \ch{the user} can define the energy function in combination with an activation condition:
\begin{equation}
    \label{eq:conditional_potential}
    E = \sum_{e \in \mathcal{E}} E_e,
    \;\;\;
    E_e =
    \begin{cases}
        E_e^+ \, &\text{if} \, c_e > 0\\
        0 \, & \text{if} \, c_e \leq 0,
    \end{cases}
\end{equation}
where $E_e^+$ is the energy of element $e$ and $c_e$ is the activation function.
To handle such expressions efficiently, \framework compiles the activation function separately and uses it to gather only the active elements for the evaluation.
In this way we can avoid the evaluation and assembly of zero energy contributions.

% --------------------------------------------------------------------------------
\paragraph{Fixed value summations}
%\label{sec:summation}

\ch{While most potentials are given by a single expression, } in numerical simulation it is common to have an inner loop per element over a set of constant data.
Such is the case in some FEM simulations where we have to evaluate an energy density function multiplied by integration weights at a set of fixed integration points.
In general, we can formulate this particular case abstractly as
\begin{equation}
    \label{eq:evaluation_loop_with_summation}
    E = \sum_{e \in \mathcal{E}} \sum_{k} \tilde{E} (\v R_e \v u; \mathcal{P}_e, \mathcal{P}_k)
\end{equation}
for some energy contribution $\tilde{E}$, where $\mathcal{P}_k$ are the parameters specific to the inner iteration $k$.
%In the example with integration points, $\mathcal{P}_k$ includes the $k$-th quadrature weight and point.

While it is possible to handle such energies by adding each iteration of the loop to the expression graph, this approach becomes expensive for complex expressions, even for moderate iteration counts.
To solve this problem, \framework compiles a single function for a symbolic set of inner iteration parameters and calls the function multiple times with updated inputs.
This approach scales well to complex models and discretizations making \framework well-suited for high-order FEM simulations as we demonstrate in Section~\ref{sec:higher_order_fem}.
%As an example Fig.~\ref{listing:fem_tet10} in Appendix~\ref{appendix:code_snippets} shows the symbolic definition of a numerical integration for a quadratic tet element with four integration points.

% --------------------------------------------------------------------------------
\paragraph{Caching compiled functions}

Symbolic differentiation, code generation and compilation typically takes less than a couple of seconds for most common expressions, as we later show in Section~\ref{sec:compilation_times}.
However, some expressions such as high-order FEM elements can take significantly longer.
To avoid unnecessary work before running simulations, \framework only differentiates and compiles new expressions or modified ones.
Compiled functions which correspond to expressions which have not changed are directly loaded.
This is achieved by storing a SHA256 hash for each energy, \ch{generated from string representations of all the expressions in the graph}, and storing it in the compiled objects.

% --------------------------------------------------------------------------------
\paragraph{Projection to positive semi-definiteness}
Our system offers optional numerical projection of element Hessian matrices to positive semi-definiteness before assembly, a common practice in second-order minimization methods to assist with convergence in Newton's method.

% --------------------------------------------------------------------------------
\paragraph{External contributions}
\label{sec:external_contributions}
Contributions to the global energy and its derivatives can also be added directly, circumventing the need for defining energy expressions.
This enhances the usability of \framework, enabling the integration of potentially faster or more robust hand-tuned derivatives when required.
% Additionally, external contributions make it possible to use energies for which symbolic differentiation solutions are unavailable and numerical methods must be used instead, such as when differentiating polar decompositions~\cite{MZS+11,CPSS10,barbic2012}.
% In such scenarios, users can provide numerical solutions for the corresponding derivatives, while still benefiting from \framework's automation for the rest of the energy sources.
\ch{
    Additionally, external contributions make it possible to use energies that require numerical approximations~\cite{MZS+11,CPSS10,barbic2012} or closed-form derivatives that need intricate procedures to be obtained and cannot be obtained by direct scalar-based differentiation~\cite{ARAP}.
}

\subsection{Assembly}
\label{sec:assembly}

In this section we describe the assembly for the general case of having multiple sets of degrees of freedom $\v u_0, \dots \v u_n$, e.g., one set for deformable volumetric solids, one for cloth models, and one for the rigid body system.

In \framework, all sets are internally concatenated into a global vector $\v u$.
This establishes a global indexing of the degrees of freedom, which is automatically considered by \framework during the assembly step.
The linear system associated with a Newton iteration then takes the form
\begin{equation}
    \def\arraystretch{1.3}
    \begin{pmatrix}
    \frac{\partial^2 E}{\partial \v u_1^2} & 
    \frac{\partial^2 E}{\partial \v u_1 \partial \v u_2} & 
    \dots & 
    \frac{\partial^2 E}{\partial \v u_1 \partial \v u_n} \\
    
    \frac{\partial^2 E}{\partial \v u_2 \partial \v u_1} & 
    \frac{\partial^2 E}{\partial \v u_2^2} & 
    \dots & 
    \frac{\partial^2 E}{\partial \v u_2 \partial \v u_n} \\
    
    \vdots  & \vdots  & \ddots & \vdots \\
    
    \frac{\partial^2 E}{\partial \v u_n \partial \v u_1} & 
    \frac{\partial^2 E}{\partial \v u_n \partial \v u_2} & 
    \dots & 
    \frac{\partial^2 E}{\partial \v u_n^2} \\
    \end{pmatrix}
    \cdot
    \begin{pmatrix}
    \Delta \v u_1 \\
    \Delta \v u_2 \\
    \vdots \\
    \Delta \v u_n \\
    \end{pmatrix}
    = -
    \begin{pmatrix}
    \frac{\partial E}{\partial \v u_1} \\
    \frac{\partial E}{\partial \v u_2} \\
    \vdots \\
    \frac{\partial E}{\partial \v u_n}
    \end{pmatrix},
    \label{eq:global_system}
\end{equation}
where $E$ is the global energy of the simulation.
The diagonal blocks in the global Hessian matrix contain the second derivatives of internal energies to a physical system, such as strain energies for deformable objects, while off-diagonal blocks contain the second derivatives of cross-system interactions, such as collisions or attachments.

\framework has default custom parallel data structures to build and return the global gradient and Hessian.
The sparse matrix structure in specific, is based on the Blocked Compressed Row Storage (BCRS) format and uses $3\times3$ matrix blocks for 3D problems.
However, our framework can also return the local element gradients and Hessians together with their global indices so that existing simulation systems can use their own data structures.

\ch{
    As discussed, contacts and remeshing, among other things, can change the sparsity pattern of the Hessian matrix between evaluations.
    However, these changes exhibit strong time coherence, meaning few non-zero elements appear or disappear from one iteration to the next (even if significant changes accumulate over a longer timescale).
    To efficiently manage these time-coherent, dynamic topology changes, the default BCRS structure in \framework adopts a dual storage strategy.
    Algorithm~\ref{algo:assembly} provides a high-level overview.
}

\ch{
    The first is a standard BCRS sparse matrix with all the values and offsets allocated contiguously in memory for high performance.
    The second is a dynamic list of ``buckets'', with one bucket per block-row, that is initialized empty at the beginning of each \chn{execution} of the assembly with the current number of block-rows.
    To insert a new block, the algorithm checks whether there is a non-zero block in the corresponding position in the BCRS matrix.
    If so, the block is simply added, avoiding expensive dynamic memory allocations.
    Otherwise, the block is appended to the corresponding block-row bucket in the second structure.
    Block insertions, regardless of whether they are added or appended, are performed in parallel using mutexes for thread synchronization.
}

\ch{
    After the insertion phase, if new blocks have been added to the buckets or any existing blocks have been left zero in the matrix, the BCRS is rebuilt, which can be efficiently done in parallel.
    This design keeps the most frequent task --- adding blocks to existing non-zero positions --- very efficient, with relatively little overhead for the much rarer changes in the non-zero structure.
}

% % ch{
%     Contacts might change the sparsity of the sparse Hessian matrix between Newton iterations.
%     This has a strong time coherency though, as not many non-zeros will appear or disappear between iterations, even though they might significantly do over time.
%     To efficiently address time-coherent dynamic topology changes in parallel, the default BCRS structure provided in \framework uses a dual storage system operated with mutexes for parallelism.
%     Algorithm~\ref{algo:assembly} show a high level overview.
%     The first structure is a standard BCRS sparse matrix and the second is a dynamic list of dynamic buckets, one per block row.
%     When a new block must be inserted, we check if the target location in the BCRS structure has been initialized and, if that is the case, we simply add it.
%     This is the preferred solution as dynamic memory allocations are avoided.
%     On the other hand, if the memory is not yet allocated for that spot (new non-zero insertion), the block is instead appended to the bucket corresponding to that block row in the second structure.
%     At the end of the insertion stage, if there are new blocks in the buckets or blocks in the matrix that are now zero, the BCRS is rebuilt, which can be efficiently done in parallel.
%     With this design, the most common task by far --- insertions to existing non-zero spots --- is kept very efficient and largely unperturbed by the few, in comparison, non-zero changes.
% % }

\begin{algorithm}
\caption{\ch{Parallel global Hessian $\v H$ evaluation and assembly. To avoid data races during parallel execution, the function \texttt{AddInPlace} and \texttt{AppendToBucketList} implement mutexes.}}
\label{algo:assembly}

$\v H \leftarrow \v 0$\tcp*[r]{Previous BCRS matrix zeroed}
\text{ClearAndResizeBucketList}($B$)\;

\ForEach{energy $E$}{
    \ForEachPar{element $e$}{
        $\v d_e \leftarrow \text{GatherData}(\text{user data, symbol-data maps})$\;
        $\v H_e \leftarrow \text{hess}_E(\v d_e)$\tcp*[r]{Call compiled function}

        \If{project}{
            \text{ProjectToPD}($\v H_e$)\;
        }

        \ForEach{block-row $i$ \textbf{\textup{in}} $\v H_e$}{
            $I \leftarrow \text{GlobalIndex}(\text{symbol-data maps}, i)$\;
            \ForEach{block-column $j$ \textbf{\textup{in}} $\v H_e$}{
                $J \leftarrow \text{GlobalIndex}(\text{symbol-data maps}, j)$\;
                \If{ \textup{BlockExists(}$\v H, I, J$\textup{)} }{
                    \text{AddInPlace}($\v H, I, J, \v H_e, i, j$)\;
                }
                \Else{
                    \text{AppendToBucketList}($B, I, J, \v H_e, i, j$)\;
                }
            }
        }
    }

    \If{\textup{HasSparsityChanged(}$\v H, B$\textup{)}}{
        $\v H \leftarrow \text{Rebuild}(\v H, B)$\;
    }
}
\end{algorithm}

\section{Applications}
\label{sec:applications}

In the following we show how complex problems in the area of physically-based simulation can be solved using our framework \framework.
\chf{
    All of the following application examples were implemented with Backward Euler time integration without loss of generality.
    As discussed in Section~\ref{sec:method_problem_definition}, other time integration methods can be formulated as an optimization problem, and could be implemented with \framework as well.
}
\ch{
    Note that we do not present an exhaustive list of all that can be accomplished with \framework; but rather, a showcase of use cases for which our system can be effectively employed.
}

\subsection{Non-Linear Material Models} \label{sec:nonlinear_materials}

\begin{figure}[tb]
    \centering
    \includegraphics[width=\columnwidth,trim={0 150 0 70},clip]{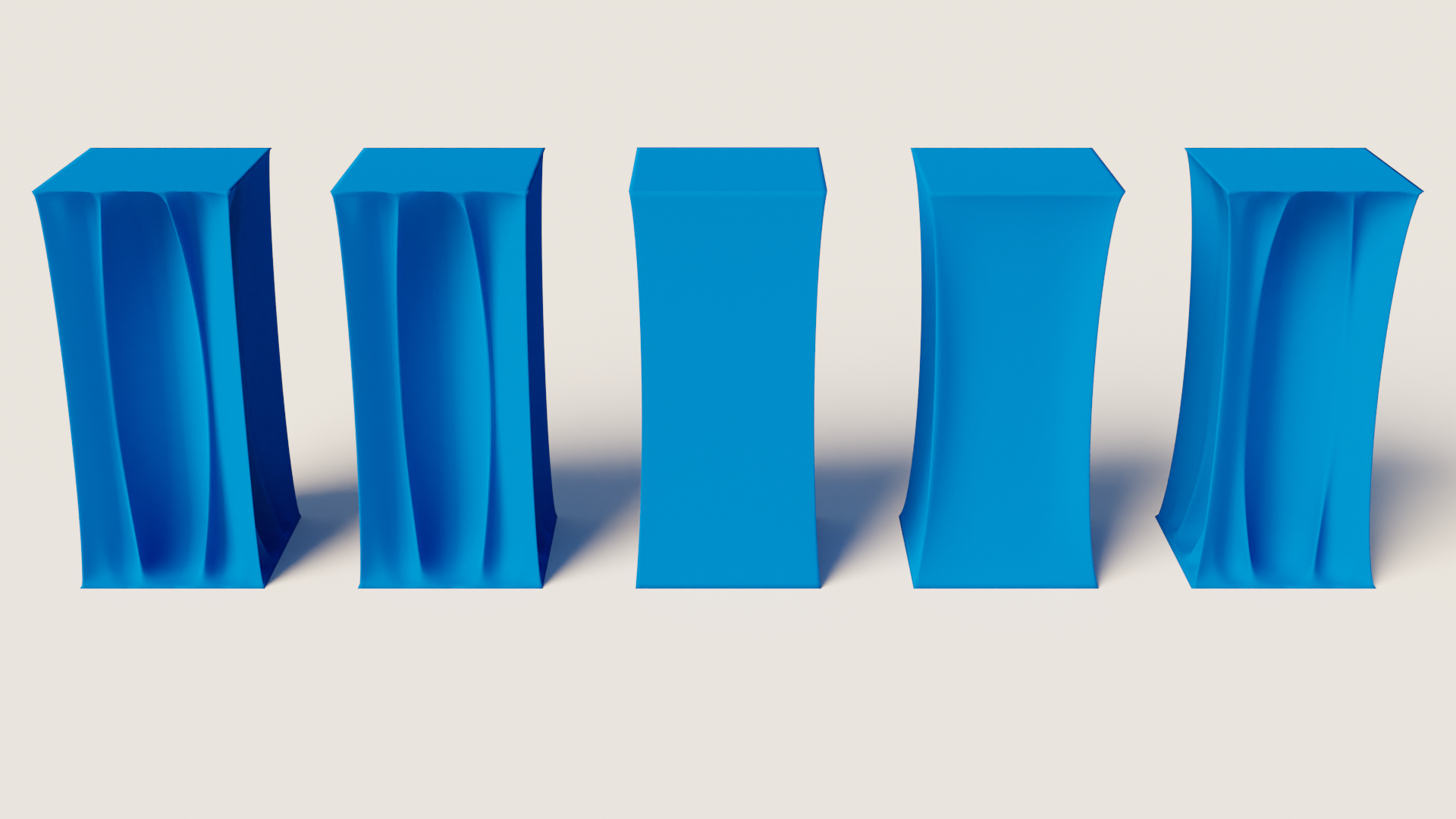}
    \caption{
    Comparison of different material models in a simulation of a stretched deformable cube.
    From left to right: ARAP with a volume conserving term, fixed co-rotational, St.~Venant-Kirchhoff, Neo-Hookean and Stable Neo-Hookean material.
    Note that we use \emph{lagged} \ch{(constant per time step)}, rotation matrices for the ARAP and fixed co-rotational energies.}
    \label{fig:material_comparison_linear}
\end{figure}

To demonstrate how concise yet powerful \framework's symbolic representation is, we implemented five different material models which took just 46 lines of code in total, see Appendix~\ref{appendix:code_snippets}.
These constitutive models are relatively complex and usually would require involved processing in the form of differentiation with respect to the deformation gradient and careful application of the chain rule.
In \framework however, we can directly use the energy expression for a given element type and let the framework work out the rest.
Fig.~\ref{fig:material_comparison_linear} presents a comparison of the implemented materials in a simulation of a stretched cube with a Young's modulus $E=$~\SI{1e4}{\pascal} and Poisson ratio $\nu=$~\SI{0.3}, showcasing the distinctive deformation behavior of such models.
\ch{
    We use lagged rotations (constant per time step)}, a long-standing common practice in computer graphics~\cite{MG04,KKB2018}, \chn{to implement the As-Rigid-As-Possible (ARAP)~\cite{SA7} and fixed co-rotational~\cite{FixedCorot} materials} in this example.
\chn{Note that such lagging introduces additional dissipation depending on the time step size~\cite[Ch.~2.5.1]{San14}.}
\ch{We also add a volume conservation term to the ARAP material~\cite{ARAP} \chn{(see also~\cite{FixedCorot})}.}
% While this is an inherent limitation of any symbolic engine, \framework makes it possible to incorporate exact rotations, if necessary, by calculating the required potentials and their derivatives with numerical methods ---in this case with the method by~Barbi\oldv{c}~\shortcite{barbic2012} for polar decompositions--- and adding them as external contributions.

\subsection{High-order \ch{Lagrangian} Finite Elements}
\label{sec:higher_order_fem}

To evaluate the energy of high-order \ch{Lagrangian} finite elements, numerical integration is typically applied using quadrature rules.
The total deformation energy of an element is then given by
\begin{equation}
    \label{eq:fem_final}
    E_e^{FEM} = \sum_{i = 1}^p w_i \, \det(\v J^e_0) \, \psi\left(\v F(\v \xi_i, \v X^e, \v x^e)\right),
\end{equation}
where $\psi$ is the strain energy density function,
$p$ is the number of integration points, and $w_i$ represents the quadrature weight.
The integration point $\v \xi_i$ is defined in the coordinate system of the reference element, and $\v J^e_0 = \v J^e_0(\v \xi_i)$ is the Jacobian of the mapping from the reference element to the physical element in the undeformed configuration (see, e.g., Wriggers et al.~\shortcite{Wriggers2008}).
The following code shows the implementation of a generic FEM integrator (Eq.~\eqref{eq:fem_final}) in \framework:
\begin{mycodeblock}{}
  Scalar fem_integration(Energy& E, 
  std::vector<Vector>& Xe, std::vector<Vector>& xe, 
  std::vector<std::array<double, 4>>& integration_points,
  std::function<Matrix(std::vector<Vector>&,Vector&)> jac,
  std::function<Scalar(Matrix& F)> psi)
  {
     Scalar sum = E.add_for_each(integration_points,
        [&](Vector& ip)
        {
           Scalar w = ip[0];
           Vector xi = Vector({ip[1], ip[2], ip[3]});
           Matrix Dm = jac(Xe, xi);
           Matrix Ds = jac(xe, xi);
           Matrix F = Ds*Dm.inv();
           return psi(F)*w*Dm.det();
        }
     );
     return sum;
  }
\end{mycodeblock}
where \texttt{jac} and \texttt{psi} are generic element Jacobian and potential energy density functions.
\ch{
   \framework includes common element Jacobians and strain energy density functions by default.
   Appendix~\ref{appendix:jacobians} shows how to use \framework to compute Jacobians, including the example of three common FEM elements: linear and quadratic tetrahedra and bilinear hexahedron.
}

\begin{figure}[tb]
    \centering
	    \includegraphics[width=\columnwidth,trim={160 180 160 20},clip]{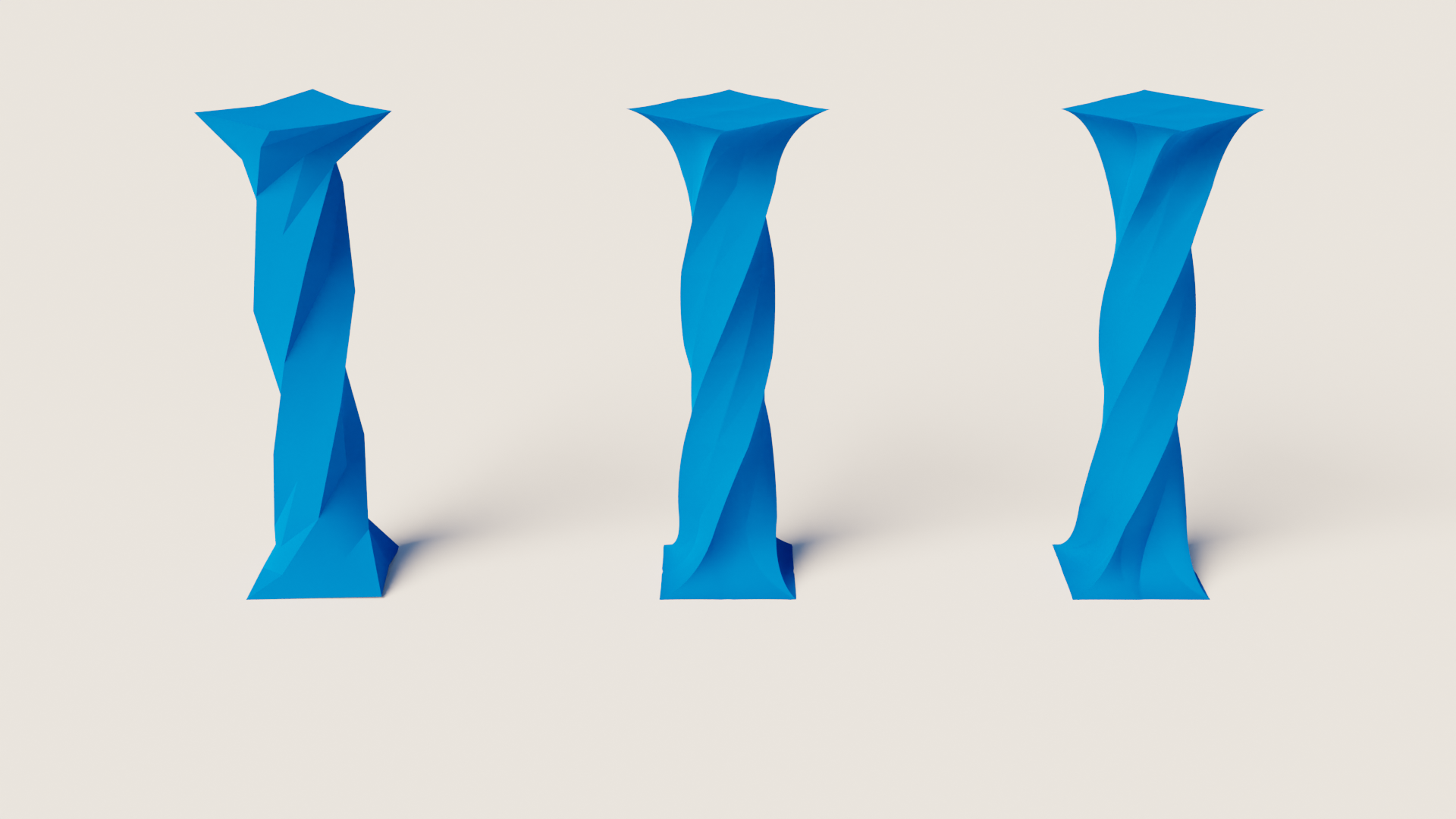}
    \caption{Comparison of linear (left), quadratic (center) and cubic (right) finite elements in a simulation of a stretched and twisted cube. }
    \label{fig:order_comparison}
\end{figure}

%In this scene, we stretch and twist a same block as in~\ref{sec:stretch_cube} but with $E = \SI{0.2}{\giga\pascal}$ until its height is $\meters{3.5}$. 
Since the quadrature points and weights are typically constant per element type, we can employ the fixed summation feature of \framework (see Section~\ref{sec:extensions}).
As a result, only the evaluation of $\psi$ at a generic integration point needs to be differentiated and compiled.
This example showcases how our framework allows for complex concepts to be expressed very concisely while preserving generality, which significantly boost productivity, reduces the room for error and ease communication between researchers.
Additionally, such high level implementations with \framework do not degrade simulation performance since the code that describes the expressions is only executed once to generate the optimized code that is actually evaluated at runtime.
Fig.~\ref{fig:order_comparison} shows a comparison of linear, quadratic, and cubic finite elements for the Stable Neo-Hookean material~\cite{StableNeoHookean}.
%An example of the Jacobian function for a tet10 element using \framework can be found in Fig.~\ref{listing:jacobian_tet10}.
%A list of constitutive models from continuous materials can be found in Fig.~\ref{listing:def_material_models}.

\subsection{Adaptive Cloth Simulation}
\label{sec:adaptive_cloth}

\begin{figure}[tb]
    \centering
    \includegraphics[width=\columnwidth,trim={150 50 250 0},clip]{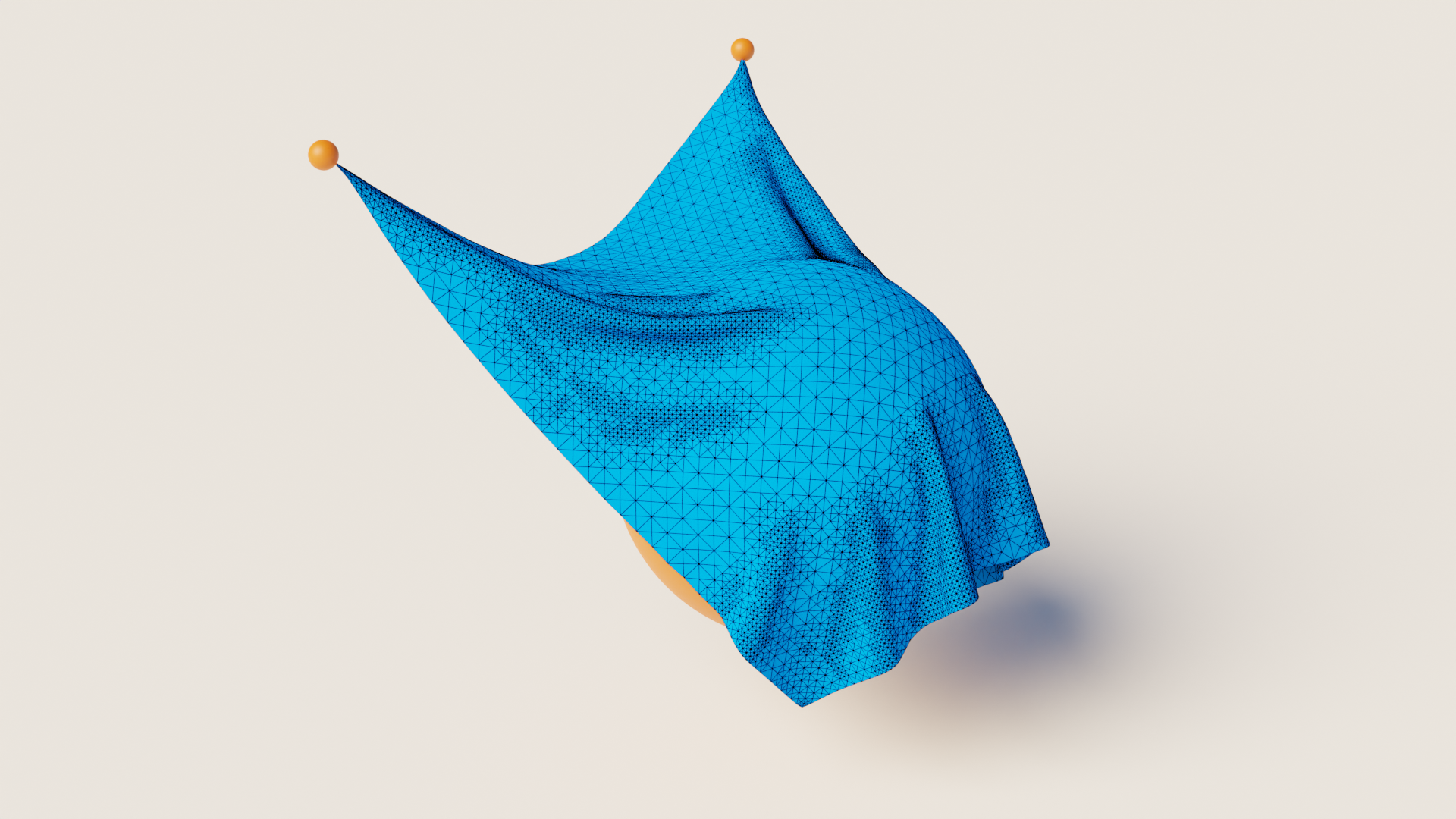}
    \caption{
    Our pipeline seamlessly handles changes in discretization, number of degrees of freedom and sparsity pattern in a cloth simulation with adaptive mesh refinement.}
    \Description{Adaptive mesh refinement cloth simulation.}
    \label{fig:refined_cloth}
\end{figure}

We implement a cloth simulation using a non-linear material in combination with a quadratic bending model, strain limiting and Rayleigh damping, in which we used an adaptive mesh refinement strategy to demonstrate that \framework can handle changes in discretization topology (see Fig.~\ref{fig:refined_cloth}).

We use the Neo-Hookean strain energy for the cloth (using a 2D FEM integrator) and the quadratic bending energy proposed by Bergou et al.~\shortcite{Bergou2006}
\begin{equation}
	E_b(\v x_{e}) = \frac{k_b}{2} \v x_{e}^T \v Q_e \v x_{e},
\end{equation}
where $k_b$ is a stiffness coefficient, $\v x_{e} \in \Real^{12}$ are the four unique mesh vertices of two adjacent triangles sharing a common internal edge $e$, and $\v Q_e \in \Real^{12 \times 12}$ is the internal edge quadratic form, which is constant during the simulation.
Implementing this energy in our system requires the precomputation of the constant matrices $\v Q_e$ and just one line of code for the energy:
\begin{mycodeblock}{}
  Scalar cloth_bending(Vector& x_e, Matrix& Q_e, 
      Scalar& k_b)
  {
     return 0.5 * k_b * x_e.transpose() * Q_e * x_e;
  }
\end{mycodeblock}

We employ a strain limiting model inspired by the one proposed by Li et al.~\shortcite{LKJ21}, where the two eigenvalues of the Green-Lagrange strain tensor $\v E = \frac{1}{2}\left( \v F^T \v F - \v I \right)$ are used to measure the strain of a triangle.
We use a simple cubic penalty with user-defined stiffness $k_{sl}$ to enforce the constraint using a $C^2$ potential energy:
\begin{equation}
   \label{eq:strain_limiting}
    E_{sl}(\v E) = \sum_i^2
    \begin{cases}
        k_{sl} A_e (\sigma_i(\v E) - \sigma_l)^3 \quad &\text{if} \quad \sigma_i(\v E) > \sigma_l\\
        0 \quad &\text{if} \quad \sigma_i(\v E) \leq \sigma_l,
    \end{cases}
\end{equation}
where $A_e$ is the undeformed area of the triangular element, $\sigma_i$ is the $i$th eigenvalue of $\v E$ and $\sigma_l$ is the user-defined stretch limiting threshold.
The implementation in our system is:
\begin{mycodeblock}{}
  Scalar cloth_strain_limiting(Matrix& F, Scalar& area, 
     Scalar& sl, Scalar& k)
  {
     Vector s = singular_value_2x2(F);
     Vector c = s - sl;
     Scalar e0 = branch(c[0] > 0, area*k*c[0].powN(3), 0);
     Scalar e1 = branch(c[1] > 0, area*k*c[1].powN(3), 0);
     return e0 + e1;
  }
   \end{mycodeblock}
The singular value decomposition of a $2 \times 2$ matrix can be computed using the direct method presented by Blinn~\shortcite{Blinn96}.

Finally, for the adaptive mesh refinement we use a quadtree subdivision scheme that splits cells based on the divergence of the normals of the mesh vertices within the quadtree node.
Although our refinement algorithm is rather simple, it suffices to show that \framework is capable of handling changes in the number of elements and degrees of freedom.
%Furthermore, this simulation also shows that the included sparse matrix assembly can handle changes in the sparsity pattern.

\subsection{Contact and Friction}
\label{sec:contact_friction}

Contact handling with friction is an important part in the simulation of deformable solids and rigid bodies and it is often a great source of complexity of the simulation model and the simulation software.
Recently, Li et al.~\shortcite{IPC} introduced the Incremental Potential Contact (IPC) method which is a robust approach to handle contact with friction.
In this section we show how the contact barrier and the friction potentials can be implemented in our framework.

First, we define a contact potential energy as
\begin{equation}
  E_{c}(d) = -k_{c}(d - \hat{d})^2\text{ln}(d/\hat{d})
\end{equation}
where $k_c$ is the barrier stiffness, $d$ the unsigned distance to the contact surface and $\hat{d}$ the maximum influence distance of the collision barrier force.
%Note that the case $d \leq 0$ is not handled by the IPC model since the method guarantees an intersection-free state.
The corresponding code in \framework is
\begin{mycodeblock}{}
  Scalar contact(Scalar& k_c, Scalar& d, Scalar& dh)
  {
     return -k_c*(d - dh).powN(2)*ln(d/dh);
  }
\end{mycodeblock}

Second, we derive the following potential energy from the IPC friction model
\begin{equation}
  E_{f}(y) = \mu f_n
  \begin{cases}
      -\frac{y^3}{3\hat{y}^2} + \frac{y^2}{\hat{y}} + \frac{\hat{y}}{3} \; &\text{if} \; y \leq \hat{y} \\
			      y \; &\text{if} \; y > \hat{y},
  \end{cases}
\end{equation}
where $y = \bignorm{\v T \Delta \v v}_2$ is the sliding contact velocity with the contact projection matrix $\v T \in \Real^{2 \times 3}$ and the relative velocity $\Delta \v v = \v v_a - \v v_b$ between the contact points $a$ and $b$.
$\hat{y}$ is the slide/stick velocity threshold, $f_n$ is the contact pressure and $\mu$ is the Coulomb's friction coefficient.
This energy is implemented in \framework as
\begin{mycodeblock}{}
  Scalar friction(Vector& va, Vector& vb, Matrix& T, 
     Scalar& mu, Scalar& fn, Scalar& yh)
  {
     Vector yt = T*(va - vb);
     Scalar y = yt.stable_norm(EPS);
     Scalar f = branch(y > yh, y, 
                   -y*y*y/(3*yh*yh) + y*y/yh + yh/3);
     return mu*fn*f;
  }
\end{mycodeblock}

Note that we must use a stable norm function that forces the returned value to be zero when $y < \varepsilon$, which is $10^{-14}\SI{}{\meter/\second}$ in our experiments, to avoid evaluating the function at a singularity which would trigger a division by zero in the derivatives.
This issue, which also cannot be circumvented using other symbolic tools like SymPy \cite{sympy}, can be avoided when the derivatives are determined by hand due to mathematical simplification.
Using stable norm works well in practice, however. %, note that the interface of \framework allows the user to integrate their own code to evaluate derivatives when a better solution is available.
Further discussion about this limitation can be found in Section~\ref{sec:limitations}.

\ch{
  In these examples we update the primitive contact pairs (point-point, point-edge, point-triangle and edge-edge) using collision detection before each energy evaluation.
  We employ an octree acceleration structure to efficiently find which primitives are in contact.
  The lists of pairs are rebuilt based on the current state of the simulation and can work with potential mesh refinement.
  As previously discussed in Section~\ref{sec:dynamic_topology}, \framework will evaluate and assemble every element of every energy contained in their respective connectivity array at the time of the evaluation call.
  Therefore, we only need to run the collision detection and update the list of pairs before requesting the global derivatives to \framework to get the correct assembly.
  }

\begin{figure}[tb]
    \centering
    \includegraphics[width=\columnwidth,trim={200 100 200 100},clip]{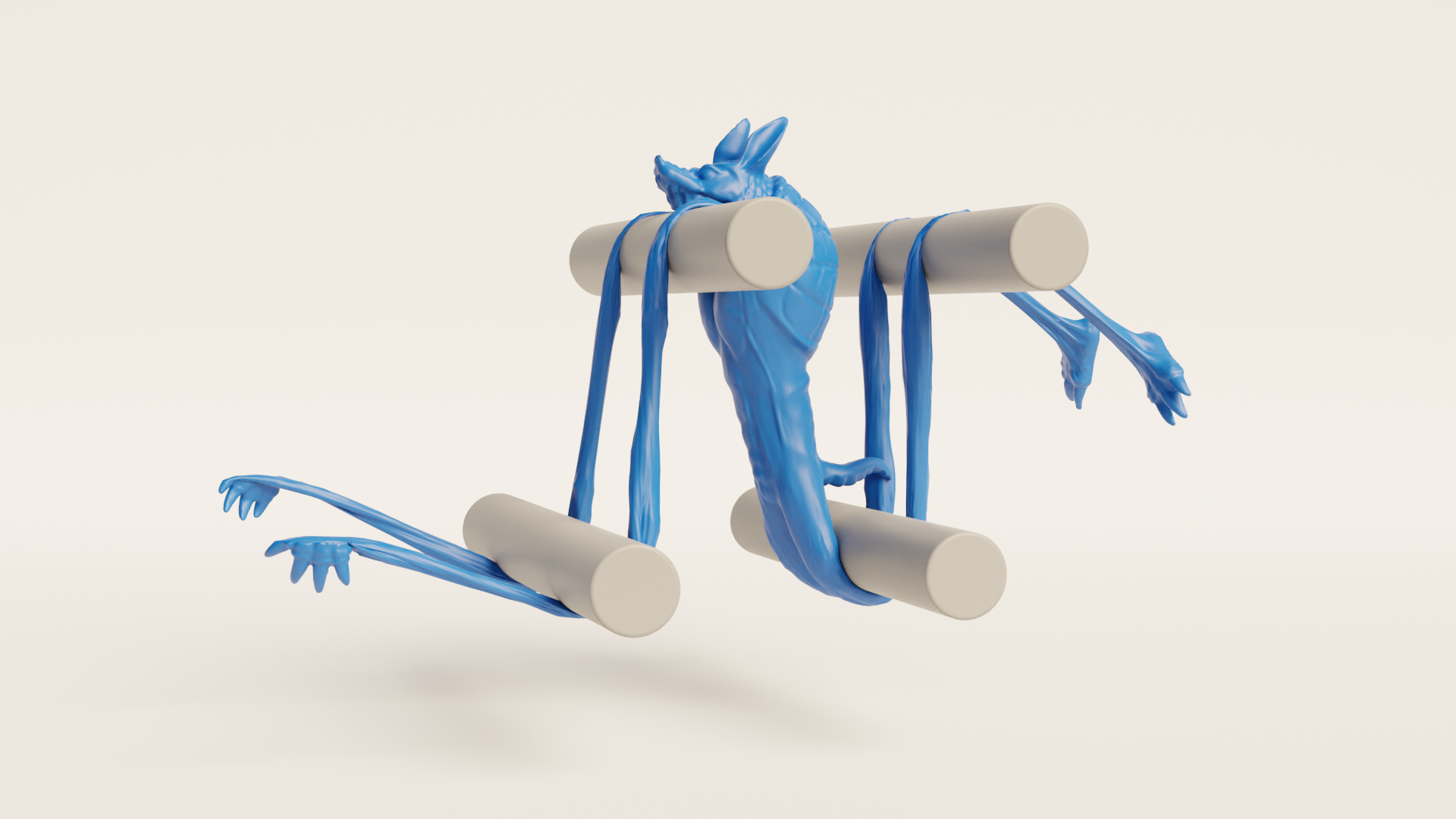}
    \caption{
		Robust contact handling in a simulation of an armadillo which is extremely deformed by animated cylinders.}
    \label{fig:armadillo}
\end{figure}

\begin{figure}[tb]
    \centering
    \includegraphics[width=\columnwidth,trim={200 0 200 0},clip]{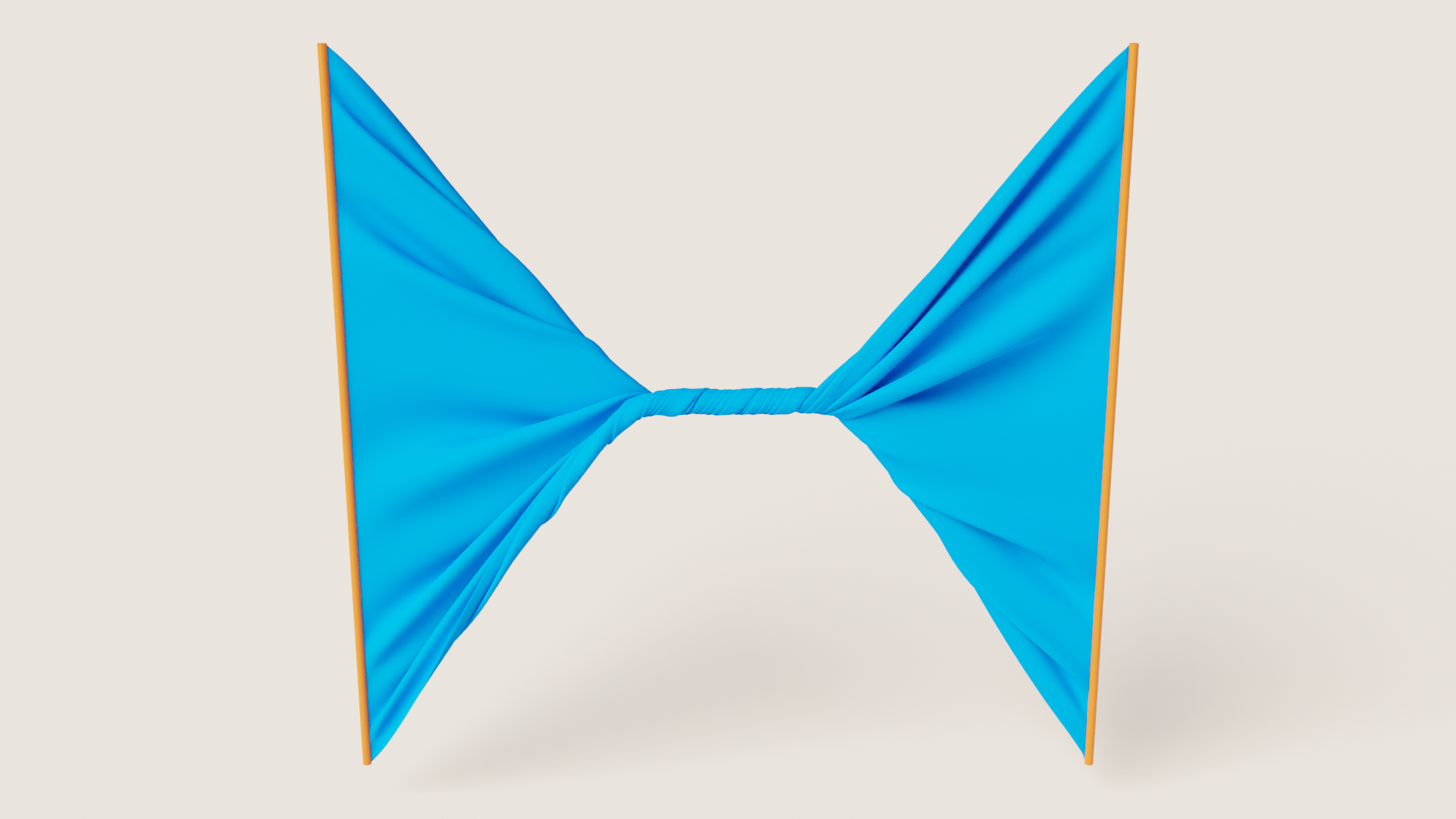}
    \caption{
		A cloth is twisted which leads to a configuration with thousands self-collisions.}
    \label{fig:cloth_wrap}
\end{figure}

We use two experiments with extreme contact configurations (see Fig.~\ref{fig:armadillo},~\ref{fig:cloth_wrap}) to show that the system enables robust evaluation of complex contact energies.
Point contacts between a deformable mesh and the environment are shown in the armadillo simulation while triangle mesh self-collisions are shown in the twisted cloth simulation which features energies based on triangle-point and mollified edge-edge distance kernels as described by Li et al.~\shortcite{IPC}.

\subsection{Coupling Multiple Systems}
\label{sec:coupling_multiple_systems}

An important feature of our proposed system is the ability to handle multiple sets of degrees of freedom, a requirement when simulating coupling between different physical systems.
%Energies that are simultaneously dependent on different sets of degrees of freedom are supported.

\paragraph*{Drifting car} %
We simulate a car model (see Fig.~\ref{fig:teaser}) by coupling a rigid body system with joints and deformable volumetric solids for the tires. 
\chf{
	For the rigid body dynamics and the corresponding inertia terms we use the formulation presented by~\citet{MEM+20}.
	Alternatively, \framework also supports implementing the potentials for the rigid body formulation introduced by~\citet{RigidIPC} or for Affine Bodies~\cite{AffineBodies}.
	Note that \framework supports the non-linear DoF mappings that typically arise in rigid body simulations (e.g. quaternion manipulations), as these can be incorporated into the energy definitions themselves.
}

We implement constraint energies using the penalty method
\begin{equation}
	E_{C} = \frac{1}{2} k_C C^2,
\end{equation}
where $k_C$ is the penalty stiffness and $C$ the constraint function.
For two connector points $a$ and $b$ with global positions $\v x_a$, $\v x_b$, velocities $\v v_a$, $\v v_b$ and two normalized direction vectors $\v d_a$, $\v d_b$, we define ball joints, direction lock constraints, and slider joints as
\begin{align}
		C^2_{bj}(\v x_a, \v x_b) &= \bignorm{\v x_a - \v x_b}_2^2 \\
		C^2_{dl}(\v d_a, \v d_b) &= \bignorm{\v d_a - \v d_b}_2^2 \\
		C^2_{sj}(\v x_a, \v x_b, \v d_a) &= \bignorm{\v x_b - \v x_a - \left((\v x_b - \v x_a)^T \v d_a \right) \v d_a}_2^2.
\end{align}
Hinge joints are simply modeled by two ball joints.
Additionally, the dampers of the car are implemented using the energy function of a damped spring
\begin{equation}
\begin{split}
	E_{ds}(\v x_a, \v x_b, \v v_a, \v v_b) =  &\frac{k_{sp}}{2} \left(\frac{\bignorm{\v x_a - \v x_b}_2}{l_0} - 1 \right)^2  +  \\
	&\frac{\alpha_{dp}}{2 l_0} \left(\left(\v v_a - \v v_b\right) \cdot \frac{\v x_a - \v x_b}{\|\v x_a - \v x_b\|_2} \right)^2,
\end{split}
\end{equation}
where $k_{sp}$ is the stiffness of the spring, $l_0$ its the rest length, and $\alpha_{dp}$ the damping coefficient.

Each wheel of the car has its own suspension system composed of multiple energies. 
A slider in combination with a damped spring models the damper of the car and attaches a rigid body to the chassis which is then linked by a hinge joint to the wheel rim to enable spinning.
We use an additional hinge joint for each front wheel to steer the car. 
Unwanted relative rotations around the slider axes are eliminated by direction lock constraints, which are also used to steer the car.
The tires are modeled by linear tet elements using the Stable Neo-Hookean material by Smith et al.~\shortcite{StableNeoHookean} and connected to the rims by attaching contacting mesh vertices with constraints analogous to ball joints.
Finally, contact and friction between the tires and the floor and obstacles are handled using the formulation introduced in Section~\ref{sec:contact_friction}.
To simplify collisions, only point-plane contacts between tires and floor are considered in this experiment.

\paragraph*{Tumble dryer} %
In the second experiment we simulate a tumble dryer with eight pieces of cloth inside.
The drum is attached to the machine's mainframe, which is fixed, by a hinge joint while torque is applied along the rotation axis.
The cloth, contact and friction models are kept as described above, the latter two are extended for coupling between cloth and rigid bodies.
This is the most complex simulation we present in this paper in regards to number of distinct energies with a total of 46, most of them being contact and friction potentials between the primitive geometries of the rigid bodies, cloth and their cross interactions.
More multi-system experiments can be found in the accompanying supplemental video.

%In the second experiment we simulated a trebuchet (see Fig.~\ref{fig:trebuchet}) by coupling a rigid body system for the trebuchet, a volumetric deformable solid system for the bunny and a rod system for the chain.
%It uses a mechanical compounded leverage to throw a soft bunny a long distance.
%To model the trebuchet we used ball joints and added a simple physical system for the string made of extensible segments.
%The bunny consists of linear finite elements in combination with the Stable Neo-Hookean material.
%Contacts are resolved with friction using the energies in Section~\ref{sec:contact_friction}.

%Fixed points to the world are modeled as ball joints where one of the points is constant.
% \input{sections/complex_materials.tex}

\section{Benchmarks}
\label{sec:results}

In the first part of this section, we present benchmarks to compare \framework, SymPy~\cite{sympy}, TinyAD~\cite{TinyAD}, and an optimized manual implementation of the Stable Neo-Hookean energy~\cite{StableNeoHookean}.
%The implementation time using SymPy took significantly longer than with the fully-automated tools since the generated code had to be integrated by hand in the simulator.
%In the comparisons we want to investigate if one of the automatically generated source codes is as fast as the highly-optimized code.
\ch{A comparison with Simit for evaluating and assembling the Neo-Hookean energy on linear tetrahedral meshes follows.}
In the end, differentiation, compilation and evaluation timings and other measurements are presented for all the simulations shown in the previous section.
Element projections to positive semi-definiteness were disabled for all experiments, as performing them would distort the assembly runtime results by adding a very significant computational cost to all methods.
%This would further equalize them and, in the view of the results, would play an arguably unfair role in closing the performance gap between manual and automatic solutions.
To prevent Newton's method from getting stuck due to indefiniteness, time steps that were too difficult (e.g. due to too many Newton iterations or a line search not descending) were restarted and half of the time step size was used instead.
After a few successful time steps, the time step size was increased again.
While for a given simulation this increases the total number of Newton iterations and therefore executions of the global assembly, the average runtime for the derivatives evaluation and assembly, which are the metrics we are actually interested in comparing, are largely unmodified.
In any case, this correction is only triggered in scenes featuring collisions.
All simulations and benchmarks were run on a workstation equipped an AMD Ryzen Threadripper PRO 5975WX with 32 cores, $3.60$ GHz and $256$ GB of RAM.
We used version 12.2.0 of the gcc compiler. %without -ffast-math as it results in slightly inconsistent results due to sometimes giving an unfair advantage to a solution but not to other and not always (in a single case, sympy is just much faster out of nowhere).

\subsection{Single Element Benchmark}
The first benchmark is the repeated evaluation of the Stable Neo-Hookean energy~\cite{StableNeoHookean}, its gradient and its Hessian for a single linear, quadratic and cubic tetrahedral element, respectively.
Note that this is a synthetic experiment aimed to \ch{\chn{assess} the performance of evaluating the derivatives in isolation, between different approaches representing different effort requirements}.
The benchmark results are shown in Table~\ref{tab:microbenchmarks}.

In regards to evaluation times, as expected, the hand-optimized solution is the fastest in all cases, with a gap that grows as the polynomial order increases.
Both symbolic differentiation approaches, SymPy and \framework, perform in the same order of magnitude than manual and within $27$\% of each other, which is also expected as they use the same fundamental principles for differentiation.
On the other hand, TinyAD is one order of magnitude slower than the other methods in all cases.
While all other approaches result in more compact final expressions due to manual or automatic reductions and simplifications, %which result in code that can be optimized very effectively by the compiler.
evaluating derivatives with TinyAD requires traversing the operation graph and applying the chain rule at each node, carrying the gradient and Hessian along.
This in turn leads to potentially more redundant operations and less room for compiler optimizations.
While in TinyAD the full derivative information is known at all intermediate operations, the other methods optimize for the final derivatives alone.
%\framework and SymPy have similar performance as they both use the same fundamental method, thus we show that we do not add any overhead by having a fully-automated pipeline.

\ch{
    SymPy's differentiation times for the linear, quadratic and cubic functions were $\SI{20.01}{\second}$, $\SI{24.19}{\minute}$, $\SI{5.93}{\hour}$ while \framework took $\SI{1.75}{\milli\second}$, $\SI{4.19}{\milli\second}$ and $\SI{14.8}{\milli\second}$, respectively.
    For reference, compilation times were $\SI{0.432}{\second}$, $\SI{2.78}{\second}$ and $\SI{25.1}{\second}$, respectively, which completely dominates the pre-simulation phase.
    %Compilation times are similar for SymPy but, since the code is integrated into the C++ code, they are hard to benchmark precisely.
}

\ch{
    % This illustrates that, while technically possible, it is not desirable to replace \framework's differentiation engine with external calls to SymPy in this case.
    These results highlight that SymPy, while being a powerful general-purpose tool, it was never intended for handling this type of complex expressions with such high-performance demands.
    Consequently, a user experimenting with complex materials or high-order integrators will face lengthy processing times.
    %This reaffirms the fact that tailor-made solutions like \framework often provide better value for the specific use case, precisely because they are designed with a very focused scope and purpose in mind.
    In any case, the code generated by SymPy is in fact relatively close to \framework's output in terms of evaluation performance, which validates \framework differentiation capabilities.
    We experimented with SymPy's \texttt{simplify} in an attempt to further reduce the final expressions complexity in addition to the already applied common subexpression elimination.
    However, it timed out after eight hours already for the Stable Neo-Hookean linear tet potential.
}

\ch{
    Finally, although proprietary mathematical engines (e.g., Mathematica, Matlab or Maple) might potentially produce derivatives faster than SymPy, they conflict with our accessibility and distribution goals outlined in Section~\ref{sec:introduction}, since they would introduce external licenses to operate the pipeline.
    % Moreover, given \framework's performance at generating derivative code, the compilation stage is by far the bottleneck accounting for more than \todo{99\%} of the pre-simulation phase.
    % Note that the compilation overhead would be similar for all differentiation engines, further reducing the incentive to consider alternatives.
}

%|               ns/op |                op/s |    err% |     total | benchmark
%|--------------------:|--------------------:|--------:|----------:|:----------
%|              230.02 |        4,347,443.49 |    0.1% |     12.11 | `sympy stable Neo-Hookean (linear)`
%|              156.94 |        6,371,865.57 |    0.1% |     12.11 | `hand-rolled stable Neo-Hookean (linear)`
%|               48.99 |       20,413,271.62 |    0.1% |     12.10 | `hand-rolled stable Neo-Hookean (linear) (energy only)`
%|            7,205.23 |          138,787.98 |    0.1% |     11.66 | `TinyAD stable Neo-Hookean (linear)`
%|              176.83 |        5,655,183.74 |    0.1% |     12.10 | `SimWorks<double> stable Neo-Hookean (linear)`

%|            6,386.43 |          156,581.97 |    0.1% |     12.10 | `sympy stable Neo-Hookean (quadratic)`
%|          365,127.52 |            2,738.77 |    0.1% |     12.10 | `TinyAD stable Neo-Hookean (quadratic)`
%|            2,825.86 |          353,875.10 |    0.1% |     12.09 | `Hand-rolled stable Neo-Hookean (quadratic)`
%|            5,908.19 |          169,256.71 |    0.1% |     12.10 | `SimWorks<double> stable Neo-Hookean (quadratic)`

%|           49,314.60 |           20,277.97 |    0.1% |     12.10 | `sympy stable Neo-Hookean (cubic)`
%|        5,300,391.62 |              188.67 |    0.1% |     12.11 | `TinyAD stable Neo-Hookean (cubic)`
%|           26,167.21 |           38,215.77 |    0.1% |     12.10 | `Hand-rolled stable Neo-Hookean (cubic)`
%|           64,132.82 |           15,592.64 |    0.1% |     12.12 | `SimWorks<double> stable Neo-Hookean (cubic)`

\begin{table}
	\caption{Average evaluation time $t_\mathrm{eval.}$ of the stable Neo-Hookean energy, its gradient and Hessian for a single linear, quadratic and cubic tetrahedral finite element.
    Relative time with respect to \framework in parenthesis.}
    \chf{
	\begin{tabular}{l|lll}
		\hline
						        & linear										& quadratic									& cubic\\
		Method 			        & $t_\mathrm{eval.} [\SI{}{\micro\second}]$ 	& $t_\mathrm{eval.} [\SI{}{\micro\second}]$ & $t_\mathrm{eval.} [\SI{}{\micro\second}]$ \\
		\hline
		Manual		 	        & $0.16$ $(\times 0.88)$ 									    & $2.83$ $(\times 0.48)$ 									& $26.17$ $(\times 0.41)$ \\
		SymPy                   & $0.23$ $(\times 1.27)$										& $6.39$ $(\times 1.08)$ 									& $49.32$ $(\times 0.77)$ \\
		\framework          	& $0.18$ $(\times 1.00)$ 									    & $5.91$ $(\times 1.00)$ 									& $64.13$ $(\times 1.00)$ \\
		TinyAD 			        & $7.21$ $(\times 40.06)$ 									& $365.13$ $(\times 61.78)$  								& $5300.39$ $(\times 82.65)$ \\
		\hline
	\end{tabular}
    }
	\label{tab:microbenchmarks}
\end{table}

\ch{We also would like to accompany the benchmarks with the qualitative experience the different solutions provided when designing the experiment.}
\ch{While just a single user experience, it is worth reporting that} an expert took roughly a work-day to differentiate, implement, test and optimize the hand-written solution.
Using SymPy, however, it took just about an hour for an experienced user to reach the solution if we exclude the time it took for SymPy to compute the derivatives themselves.
Note that while generic symbolic tools have comprehensive differentiation and code generation modules, the exact functionality needed for this specific task was not readily available and some scripting was required.
%The implementation with SymPy required some back-and-forth between the differentiation and code generation Python scripts and the C++ codebase.
The code footprint of the SymPy solution was significantly larger than for the other three approaches due to the code being divided between scripts and main application.
%Also, incorporating the rather large generated code to the C++ build leads .
Finally, both TinyAD and \framework presented the most streamlined processes, allowing for a trained user to reach the solution in about ten minutes, including the time to differentiate and to compile the all the expressions.
This is due to both tools being fully-automated and specifically designed for this type of task.
Notably, typos and other bugs were much less problematic as the code itself is very short and corrections to the root expressions have an immediate impact, unlike the two previous methods.
% \ch{While only a single user experience, it corresponds with similar reports from Simit~\cite{simit}, which we believe naturally reflect the nature of this type of frameworks.}

\subsection{Simulation Benchmark}
In this experiment we run a benchmark in a more realistic setting where we compare the total runtime of the derivatives evaluation and assembly during a simulation. %, averaged over all Newton iterations.
%Note that for a fair comparison, we exchanged TinyAD's triplet based sparse matrix assembly with our block based assembly which is significantly faster.
We use the simulation setup shown in Fig.~\ref{fig:order_comparison} with linear, quadratic and cubic tetrahedral elements sharing the same 137K degrees of freedom, and present the timings in Table~\ref{tab:simbenchmarks}.
TinyAD and \framework use their respective built-in data structures and assembly.
Manual and SymPy use \framework's global data structures by being declared into a simulation as external contributions, see Section~\ref{sec:extensions}.

The manually optimized solution is again the fastest in all cases and TinyAD is again the slowest, with \framework being up to $361x$ faster for cubic elements.
\framework and SymPy are again roughly matched in performance, and both close the gap to the manual solution due to the more realistic execution conditions.
%memory traffic conditions and the assembly being an equal cost for all three solutions.
The implementation effort of using TinyAD and \framework is again similar, both the lowest by a significant margin, as the global solutions are generated directly from the mathematical expressions and global data structures are provided.
% \ch{For the the SymPy and manual implementations, we used \framework's data structures and assembly procedures.}
%took significantly longer to implement in comparison to the fully-automated approaches since evaluation and the integration into the assembly pipeline had to be written by hand.
Both SymPy and manual evaluations allowed for further optimizations, which made the assembly faster than using the generic evaluation and assembly procedure in \framework, and resulted in the SymPy solution being ultimately roughly 10\% faster, not by own merits, but by virtue of a hand tuned assembly.
\framework on the other hand must account for arbitrary element types, and different function inputs and outputs, information which is only made available at runtime.
% \todo{Scripting SymPy would incur the same cost...}
%On the other hand, manual evaluation and assembly can be specific for each function type and ... which leads to more effective compiler optimizations, such as inlining.
%The downside, other than higher upfront development effort, is a larger code footprint to maintain as the simulator grows.

In any case, it is important to emphasize that the difference in total simulation runtime between methods is much less pronounced than what Table~\ref{tab:simbenchmarks} might suggest, as evaluation of derivatives and assembly is just part of the total execution time.
\ch{
    The largest portion of the total simulation runtime is usually spent in the linear system solve (not included in Table~\ref{tab:simbenchmarks}), which often plays a significant role as performance equalizer.
}

\ch{
    \framework employs a $3\times3$ Block Diagonal Preconditioned Conjugate Gradient linear system solver with a forcing sequence tolerance~\cite{Nocedal}.
    In our experiments, the average runtimes for the linear solves are $\SI{29.8}{\milli\second}$, $\SI{52.9}{\milli\second}$ and $\SI{256.4}{\milli\second}$ for the linear, quadratic and cubic \framework's simulations, respectively.
    Both the manual and SymPy simulations used \framework's linear solver, while TinyAD uses its own solver.
}
Note that these times are for reference and that analyzing the role of linear system solvers in optimization time integrators is out of the scope of this work as different choices can have a large impact in overall simulation runtime.
For example, choosing between direct or iterative solver is a decision that might be conditioned by simulation size or expected numerical stiffness and which might drastically change the total simulation runtime independently of the derivative computation.
%In any case, to give a rough reference to the reader, excluding the TinyAD simulations, we use a custom Conjugate Gradient solver in this benchmark, which we measured to be \todo{X} times faster than Eigen's on average.
%In all the simulations run with \framework, the linear system solve took upwards of 90\% of the total simulation time, with even higher proportions for high-order FEM.
%Additionally, this simulation does not feature contacts, which would also play an important role further equalizing the timings.
% In the context of a full simulation, the advantage that hand-optimized manual derivatives has in the single element benchmark shown above is severely diminished, which in turn significantly enhances the value offered by our fully-automated system.
%We conclude that while both \framework and TinyAD can be effectively used for prototyping, the substantial performance gap between them makes our proposed system much more suitable as it will scale much better to larger simulations.

%CG times: 92.2%, 97.7%, 98.1%

\begin{table}
	\caption{Simulation benchmark of the stretched cube example with 137K degrees of freedom using linear, quadratic and cubic tetrahedral finite elements.
        Timings are averaged per Newton iteration and include the evaluation of the energy and its derivatives as well as the assembly.
        The linear system solve is not included in the timings.
    }
    \chf{
	\begin{tabular}{l|lll}
		\hline
						    & linear									    & quadratic									    & cubic\\
		Method 			    & $t_\mathrm{total} [\SI{}{\milli\second}]$  & $t_\mathrm{total} [\SI{}{\milli\second}]$  & $t_\mathrm{total} [\SI{}{\milli\second}]$ \\
		\hline
		Manual	            & $12.5$ $(\times 0.84)$     & $17.9$ $(\times 0.74)$         & $28.8$ $(\times 0.54)$ \\
		SymPy               & $14.3$ $(\times 0.97)$ 	& $22.2$ $(\times 0.91)$         & $46.6$ $(\times 0.88)$ \\
		\framework          & $14.8$ $(\times 1.00)$     & $24.3$ $(\times 1.00)$         & $52.9$ $(\times 1.00)$ \\
		TinyAD       	    & $698.9$ $(\times 47.22)$    & $1269.7$ $(\times 52.25)$       & $19125.2$ $(\times 361.53)$ \\
		\hline
	\end{tabular}
    }
    \label{tab:simbenchmarks}
\end{table}

\begin{figure}[tb]
    \centering
    \includegraphics[width=\columnwidth,trim={0 0 0 0},clip]{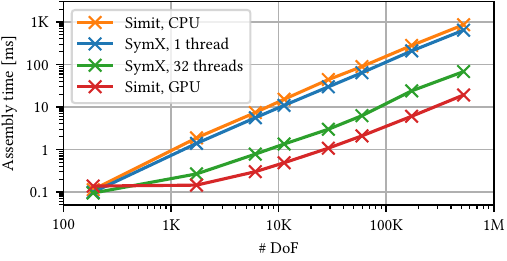}
    \caption{
		\chf{Total evaluation and assembly time of the Hessian for the dynamic simulation of a cube with Neo-Hookean material and linear tetrahedra. Simit does not parallelize assembly on the CPU. 
        }}
    \label{fig:assembly}
\end{figure}

\subsection{Comparison with Simit}
\label{sec:simit_comparison}
\chf{
    In this section we evaluate and compare the performance of \framework with Simit~\cite{simit}.
    While Simit does not offer differentiation capabilities, it does provide code generation and automates the assembly process over sets of abstract elements.
    For the experiment, we use the previously introduced stretched cube experiment with linear tetrahedral elements and the classic Neo-Hookean material model.
    We measure the average time required for the derivatives evaluation and the assembly of the global Hessian for varying mesh resolutions.
    %Note that the linear solver is a completely exchangeable component in both frameworks, therefore those timings are not included.
    The results are shown in Fig.~\ref{fig:assembly}.
}

\ch{
    Note that while \framework supports parallel evaluation and assembly (Section~\ref{sec:assembly}), Simit only provides sequential evaluation and assembly on the CPU.
    Comparing Simit and \framework in single thread execution, the evaluation and global assembly in \framework is between 15\% and 48\% faster depending on the resolution (34.7\% on average over all resolutions).
    % The smallest gain is measured for the coarsest resolution with 189 degrees of freedom (0.099ms vs. 0.113ms) while the largest resolution with 524745 degrees of freedom is 34.1\% faster in \framework (641ms vs. 860ms).
    These competitive results for \framework validate its evaluation and assembly procedures in relation to established tools.
    With multi-threading enabled, we measure speedups between 9.5 and 12.7 times in comparison to Simit if we exclude simulations with fewer than 6075 degrees of freedom.
    Fig.~\ref{fig:assembly} also includes runtimes corresponding to Simit's GPU assembly for this experiment.
    Using a NVIDIA GeForce RTX 3090 Ti, Simit reports global assembly runtimes between 2.6 and 4.0 times faster than \framework's multithreaded execution, if we exclude simulations with fewer than 6075 degrees of freedom.
}

\ch{
    It is important to note that DSLs come with challenges beyond the potentially large number of derivatives they would need to obtain (e.g. for the dryer scene of Fig.~\ref{fig:dryer}): collision detection, changing topologies, minimizer and linear solver and line search, are some of the components that need to be done either in the DSL scripting language itself or communicating with the host language and device.
}

\subsection{Application Example Timings}
In addition to these benchmarks, we also show timings of the application examples from Sec.~\ref{sec:applications} in Table~\ref{tab:sim_table}.
% Again, the time $t_\mathrm{total}$ includes the runtime just for evaluation and assembly averaged per Newton iteration.

\begin{table}
    \caption{Number of degrees of freedom and average evaluation and assembly timings per Newton iteration of the application examples from Sec.~\ref{sec:applications}.
    }
    \chf{
    \begin{tabular}{cl|cc}
    		\hline
        Fig. & Scene & \# DoF & $t_\mathrm{total} [\ms{}]$ \\
        \hline
        \hline
        \ref{fig:material_comparison_linear} & ARAP 				& $307623$ 	& $38.79$ \\
        \ref{fig:material_comparison_linear} & Fixed Corot. 		& $307623$ 	& $35.40$ \\
        \ref{fig:material_comparison_linear} & StVK 			    & $307623$ 	& $37.85$ \\
        \ref{fig:material_comparison_linear} & Neo-Hookean 			& $307623$ 	& $35.83$ \\
        \ref{fig:material_comparison_linear} & Stable NH 	    & $307623$ 	& $35.10$ \\
        \hline
        \ref{fig:order_comparison} 	 & Beam linear 		& $189$ 		& $0.09$ \\
        \ref{fig:order_comparison} 	 & Beam linear 		& $1728$ 	& $0.24$ \\
        \ref{fig:order_comparison} 	 & Beam linear 		& $11253$ 	& $1.19$ \\
        %\hline
        \ref{fig:order_comparison} 	 & Beam quadratic 	& $975$ 		& $0.24$ \\
        \ref{fig:order_comparison} 	 & Beam quadratic 	& $11253$ 	& $1.75$  \\
        \ref{fig:order_comparison} 	 & Beam quadratic 	& $80703$ 	& $14.42$ \\
        %\hline
        \ref{fig:order_comparison} 	 & Beam cubic 		& $2793$ 	    & $0.92$ \\
        \ref{fig:order_comparison} 	 & Beam cubic 		& $35328$ 	& $10.93$ \\
        \ref{fig:order_comparison} 	 & Beam cubic 		& $262353$ 	& $96.46$ \\
        \hline
        \ref{fig:refined_cloth} 	 & Adaptive cloth & $6339 - 24903$ 	& $4.19$ \\
        \hline
        \ref{fig:armadillo} & Armadillo 	& $195510$ 	& $29.56$ \\
        \hline
        \ref{fig:cloth_wrap} & Twisted Cloth 	& $482403$ 	& $95.64$ \\
        \hline
        \ref{fig:teaser} & Car 	& $17220$ 	& $2.89$ \\
        \hline
        \ref{fig:dryer} & Dryer 	& $244836$ 	& $73.33$  \\
        \hline
    \end{tabular}
    }
    \label{tab:sim_table}
\end{table}

\subsection{Compilation Times}
\label{sec:compilation_times}
Finally, we also present the timings and memory requirements associated with differentiation and compilation for all the experiments in Table~\ref{tab:comp_table}.
Here, $T_\mathrm{diff}$ denotes the total time needed for differentiating and generating the code for all the expressions in the simulation and $T_\mathrm{comp}$ the time for gcc to compile the generated C++ code.
Concerning the memory requirements, the peak memory column indicates the maximum storage needed during differentiation and the binary size is the sum of all the binaries generated for that simulation, which is shown per scene for illustration purposes.
In practice, a simulation software would keep the binaries for all the implemented potentials and only the relevant ones for a given simulation instance would be loaded.
Timings for differentiation and compilation strongly correlate to expression complexity, which explains why the beam simulation with the cubic FEM discretization takes the longest to differentiate and compile.
Our memory requirements during differentiation are very low, specially in comparison to relevant methods such as the one by Herholz et al.~\shortcite{SymbolicLib} which compiles the whole problem, needing tens of gigabytes to differentiate moderately sized simulations.
%For example, we did not have to increase the standard stack size in any of our experiments and benchmarks as it was necessary using TinyAD.
Additionally, the disk space required to store our binaries is very modest, in contrast to the gigabytes needed by Desai et al.~\shortcite{ACORNS}.
%\todo{re-run timing the actual whole differentiation, not the acc, as we have more energies than threads.}

% SymPy differentiation times for linear, quadratic and cubic with SNH material: 20.009279 s, 1451.511863 s, 21353.671351 s

\begin{table}
    \caption{Measurements regarding \framework's initialization.
    $T_\mathrm{diff}$ indicate total differentiation and code generation runtime, and $T_\mathrm{comp}$ compilation times.
    Compilation was carried out from scratch, that is, no cached function was loaded from disk.
    Note that all energies are initialized in parallel.
    Furthermore, the table shows the peak memory consumption needed during differentiation and the total size of the output binaries.}
    \chf{
    \begin{tabular}{cl|rrrr}
        \hline
        & & $T_\mathrm{diff} $ & $T_\mathrm{comp}$ & Memory & Binary \\
        Fig. & Scene & $[\ms{}]$ & $[\ms{}]$ & [kB] & [kB] \\
        \hline
        \hline
        \ref{fig:material_comparison_linear} & ARAP 				& $1.5$ 	& $675.0$ 	& $61$ 	& $182$ \\
        \ref{fig:material_comparison_linear} & Fixed Corot. 		& $1.3$ 	& $669.7$ 	& $57$ 	& $178$ \\
        \ref{fig:material_comparison_linear} & StVK 			& $1.9$ 	& $853.7$	& $115$ 	& $180$ \\
        \ref{fig:material_comparison_linear} & Neo-Hookean 			& $1.4$ 	& $752.0$ 	& $72$ 	& $182$ \\
        \ref{fig:material_comparison_linear} & Stable NH 	& $1.7$ 	& $737.8$ 	& $68$ 	& $182$ \\
        \hline
        \ref{fig:order_comparison} 	 & Beam linear 	    & $1.6$ 	& $722.9$ 	& $68$ 	& $182$ \\
        \ref{fig:order_comparison} 	 & Beam quadratic    & $12.1$ 	& $3273.3$ 	& $309$ 	& $260$ \\
        \ref{fig:order_comparison} 	 & Beam cubic 	    & $40.3$ & $26309.4$ 	& $1090$ & $498$ \\
        \hline
        \ref{fig:refined_cloth} & Adaptive cloth & $1.6$ & $768.4$ & $118$ & $332$ \\
        \hline
        \ref{fig:armadillo} & Armadillo & $4.4$ & $1395.6$ & $483$ & $775$ \\ 
        \hline
        \ref{fig:cloth_wrap} & Twisted cloth & $3$ & $1255$ & $1229$ & $873$ \\
        \hline
        \ref{fig:teaser} & Car & $6.7$ & $1655.2$ & $545$ & $911$ \\
        \hline
        \ref{fig:dryer} & Dryer & $10.3$ & $3676$ & $5889$ & $4505$ \\
        \hline
    \end{tabular}
    }
    \label{tab:comp_table}
\end{table}

\section{Limitations and Future Work}
\label{sec:limitations}

%Many state-of-the-art simulation methods use a variant of \emph{Projected Newton}~\cite{GSS+15,IPC,LKJ21} to prevent Newton's method from getting stuck due to indefiniteness of the global Hessian matrix.
%Here the element-local Hessian contributions are projected to semi-definiteness in some way.
%Our system can be used with Projected Newton by computing eigendecompositions of the element matrices.
%However, this may be more computationally expensive than state-of-the-art methods for deformable solids that rely on known properties of the eigenstructure of the material models themselves~\cite{KGI19,SGK19,StableNeoHookean}.
%We believe that the techniques we present in this paper could be enhanced with analytic eigenstructures to conveniently and efficiently differentiate material models defined in terms of scalar invariants.

%Although we demonstrate that symbolic differentiation and code generation gets much closer to manually implemented code than the alternatives, there is still room for improvement.
%Our current system does not mathematically simplify the resulting derivative expressions besides avoiding redundant computation.

Currently, our system cannot mathematically simplify expressions.
Although mathematical simplification has limited performance implications, as shown by Herholz et al.~\shortcite{SymbolicLib}, it would certainly help further reducing the gap to \chn{manually} optimized code.
%It is possible that adding this kind of capability, as done in \cite{SymbolicLib}, might help make the gap even smaller.
In a similar spirit, directly supporting vectors and matrices in the expression graph instead of eagerly reducing all quantities to scalar operations might aid the search for more compact expressions.
Finally, while also not a fundamental limitation, \framework currently only works with fixed element sizes.
Handling element connectivity with dynamic sizes is left for future work.

The treatment of hard constraints, e.g. in the form of Lagrange Multipliers, falls beyond the scope of this work.
Nevertheless, we anticipate no significant challenges in expanding \framework to handle these hard constraints and their derivatives as long as they can be symbolically represented.

\paragraph*{Analytic projections} %
Although our system can project element matrices to positive semi-definiteness, some approaches can exploit the properties of the eigenstructure of the material models themselves \cite{KGI19,SGK19,StableNeoHookean} to perform a more efficient projection.
We believe that the techniques we present in this paper could be enhanced with analytic eigenstructures to conveniently and efficiently differentiate material models defined in terms of scalar invariants.

\chf{
    \paragraph*{Tensor identities} %
    \framework currently employs a scalar-based engine for the representation and differentiation of expressions, and therefore it does not implement tensor identities or tensor differentiation rules in its current form.
    Consequently, it cannot independently find derivatives for which such identities are required.
    %This implies that, for example, \framework cannot compute the rotationally exact ARAP derivatives autonomously by means of direct scalar differentiation.
    However, it is possible to incorporate such energies in \framework as external contributions, for instance using the closed form expressions of the ARAP derivatives presented by \citet{ARAP}.
    This is a limitation of the scalar nature of the current engine --- a constraint shared with other scalar-based engines --- not of the overall concept presented.
    This limitation could be addressed in future work by extending the symbolic engine to handle tensor identities.
}

\paragraph*{Singularities} %
Arguably the most significant limitation of symbolic and automatic differentiation is that some expressions, while analytically differentiable, may contain partial expressions in their expression graph that are non-differentiable.
Therefore, evaluating the result near a non-differentiable point in the intermediate expression may cause the intermediate result to become numerically unstable.
Typically these kind of issues occur when the scalar expression contains norms, square roots or more generally fractional powers, as we have already seen in the symbolic definition of the friction energy, Section~\ref{sec:contact_friction}.
Users may be taught to be wary of these issues in the presence of such expressions and apply workarounds like smooth approximations provided by the framework, but ultimately this is not foolproof.
A mechanism for automatic reformulation of the expression to avoid such problems would be an improvement.
In the interim, the system could be augmented to optionally detect such potential problems and notify the user, so that they may try a different formulation or use the stable operators provided.

Finally, a natural next step for \framework to improve its \chn{performance} is to provide support for GPU execution.
\chn{
Since SymX already implements read-only views of the data required for computing global derivatives, synchronizing data between the host and device would be straightforward.
Extending the code generation process to produce GPU-compatible code for derivative evaluation can be accomplished analogously to the existing CPU code generation. 
Assembly and linear algebra operations could be performed using standard GPU libraries such as cuSPARSE~\cite{NVIDIA_cuSPARSE_12_9}.
}

\section{Conclusions}
In this work we presented \framework, a system to automate the differentiation and assembly in complex simulations based on optimization time integrators.
The proposed system provides a set of symbolic types that allows engineers and researchers to succinctly define the different contributions to the global energy of the simulation.
Thanks to the link between these symbols and the simulation data, the system can apply symbolic differentiation to the energies with respect to the degrees of freedom of the simulation and completely automate the evaluation and assembly process.
Thanks to on-the-fly compilation of the derivatives code, our method has a performance comparable to code optimized by hand.

We demonstrated the capabilities of our method in an array of challenging simulations featuring state-of-the-art models and showed that the code required to express such simulation energies very closely resembles their original mathematical counterparts.
In the view of the results obtained, we conclude that \framework can significantly support engineers by allowing them to quickly prototype fast and reliable simulation software with a minimal code footprint, that is also easy to understand and distribute.
Changes in the expressions are immediately incorporated in the simulations which allow researchers to experiment with new models, or variations of existing ones, and to quickly reach results.
The flexibility of our method also presents a path for an initial prototype to be gradually transitioned to a hybrid between symbolic and manual derivatives as the user sees fit.
\framework is therefore a great candidate to provide flexible and powerful symbolic facilities to higher level simulation codebases that focus on other aspects, such as different types of material discretization or time integration.
Finally, we are convinced that also other simulation methods like constraint-based approaches, or even applications in different fields like geometry processing, will benefit from our framework.

%%
%% Acknowledgements
\begin{acks}
  Fabian Löschner and Andreas Longva are funded by the Deutsche Forschungsgemeinschaft (DFG, German Research Foundation) — 281466253; 411281008.
\end{acks}

\bibliographystyle{ACM-Reference-Format}
\bibliography{symx_paper}

\appendix
\section{\ch{Jacobians}}
\label{appendix:jacobians}

\ch{
We illustrate now how to define Jacobian functions with \framework for Lagrangian FEM simulations.
The Jacobian of an element is
\begin{equation}
    \v J_{\v x} = \frac{\partial \v x}{\partial \v \xi}
\end{equation}
where $\v x$ are the coordinates in the current configuration and $\v \xi$ the coordinates in the reference configuration.
The Jacobian at rest configuration $\v J_{\v X}$ is defined analogously.
Space is typically discretized as $\v x \approx \v N(\v xi) \v x_h$ where $\v N$ are the interpolation (or shape) functions and $\v x_h$ are the coordinates of the nodes of the discretization element.
}

\ch{
The following is a generic function to compute Jacobians symbolically with \framework
}
\begin{mycodeblockscriptsize}{}
  Matrix fem_jacobian(const FEM_Element_Type& element_type, 
     const std::vector<Vector>& xh, const Vector& xi)
  {
     int N = xi.size();
     Vector v = interpolation(element_type, xh, xi);
     std::vector<Scalar> jac;
     for (int i = 0; i < N; i++) {
        for (int j = 0; j < N; j++) {
           jac.push_back(diff(v[i], xi[j]));
        }
     }
     return Matrix(jac, { N, N });
  }
\end{mycodeblockscriptsize}

\ch{
For completeness, we include below the interpolation scheme of three common FEM element types: linear and quadratic tetrahedra and bilinear hexahedron~\cite{shapefunctions}.
}
\begin{mycodeblockscriptsize}{}
  enum class FEM_Element_Type { Tet4, Tet10, Hex8 };
  
  template<typename T>
  T interpolation(const FEM_Element_Type& element_type, 
     const std::vector<T>& v, const Vector& xi)
  {
     if (element_type == FEM_Element_Type::Tet4) {
        Vector N = Vector({
           1.0 - xi[0] - xi[1] - xi[2],
           xi[0],
           xi[1],
           xi[2]});
        return dot(N, v);
     }
     else if (element_type == FEM_Element_Type::Tet10) {
        Scalar N0 = 1.0 - xi[0] - xi[1] - xi[2];
        Scalar N1 = xi[0];
        Scalar N2 = xi[1];
        Scalar N3 = xi[2];
        Vector N = Vector({
           N0 * (2.0 * N0 - 1.0),
           N1 * (2.0 * N1 - 1.0),
           N2 * (2.0 * N2 - 1.0),
           N3 * (2.0 * N3 - 1.0),
           4.0 * N0 * N1,
           4.0 * N1 * N2,
           4.0 * N2 * N0,
           4.0 * N0 * N3,
           4.0 * N1 * N3,
           4.0 * N2 * N3});
        return dot(N, v);
     }
     else if (element_type == FEM_Element_Type::Hex8) {
        Scalar NXm = 0.5 * (1.0 - xi[0]);
        Scalar NXp = 0.5 * (1.0 + xi[0]);
        Scalar NYm = 0.5 * (1.0 - xi[1]);
        Scalar NYp = 0.5 * (1.0 + xi[1]);
        Scalar NZm = 0.5 * (1.0 - xi[2]);
        Scalar NZp = 0.5 * (1.0 + xi[2]);
        Vector N = Vector({
           NXm * NYm * NZm,
           NXp * NYm * NZm,
           NXp * NYp * NZm,
           NXm * NYp * NZm,
           NXm * NYm * NZp,
           NXp * NYm * NZp,
           NXp * NYp * NZp,
           NXm * NYp * NZp });
        return dot(N, v);
     }
  }
\end{mycodeblockscriptsize}

\ch{
In Section~\ref{sec:higher_order_fem} we show a generic FEM integrator written with \framework that can take any Jacobian function \texttt{jac} with the following signature
}
\begin{mycodeblockscriptsize}{}
  std::function<Matrix(std::vector<Vector>&,Vector&)> jac;
\end{mycodeblockscriptsize}
\ch{
In order to use the example jacobian function provided in this appendix one needs to select an element before passing it to the integrator.
For example
}
\begin{mycodeblockscriptsize}{}
  auto jac = [element_type](std::vector<T>& v, Vector& xi){ 
       fem_jacobian(element_type, v, xi); 
    };
\end{mycodeblockscriptsize}

\ch{
Similar selection would need be done for the constitutive models in Appendix~\ref{appendix:code_snippets}.
Note that this flexible composability using C++ lambdas does not have performance implications at simulation time since the symbols are processed and compiled regardless of how the expressions are constructed.
}

\newpage
\section{Constitutive Models}
\label{appendix:code_snippets}
Here we show the implementation in \framework of the five constitutive models~\cite{StableNeoHookean,ARAP,FixedCorot} used in the simulation shown in Fig.~\ref{fig:material_comparison_linear}.

\begin{mycodeblockscriptsize}{}
  Scalar constitutive_models_energy_density(Energy& energy,
     Matrix& F, Matrix& R,  /* R is assumed constant*/
     Scalar& lambda, Scalar& mu, 
     ConstitutiveModels model)
  {
     // Eq. 14 from [Smith et al. 2018]
     if (model == ConstitutiveModels::StableNeoHookean) {
        Scalar mu_ = 4/3*mu;
        Scalar lambda_ = lambda + 5/6*lambda;
        
        Matrix C = F.transpose()*F;
        Scalar detF = F.det();
        Scalar Ic = C.trace();
        Scalar alpha = 1 + mu_/lambda_ - mu_/(4*lambda_);
        return 0.5*mu_*(Ic - 3) + 
           0.5*lambda_*(detF - alpha).powN(2) - 
           0.5*mu_*log(Ic + 1);
     }
     
     // Eq. 5 from [Smith et al. 2018]
     else if (model == ConstitutiveModels::NeoHookean) {
        Matrix C = F.transpose()*F;
        Scalar Ic = C.trace();
        Scalar logdetF = log(F.det());
        return 0.5*mu*(Ic - 3) - mu*logdetF + 
           0.5*lambda*logdetF.powN(2);
     }
     
     // Eq. 49 from [Smith et al. 2018, Stomakhin et al. 2012]
     else if (model == ConstitutiveModels::FixedCorot) {
        Matrix I = energy.make_identity_matrix(3);
        Scalar detF = F.det();
        return mu*(F - R).frobenius_norm_sq() + 
           0.5*lambda*(detF - 1).powN(2);
     }
     
     // Eq. 14 and 36 from [Lin et al. 2022]
     else if (model == ConstitutiveModels::ARAP) {
        Matrix C = F.transpose()*F;
        Scalar Ic = C.trace();
        Scalar detF = F.det();
        return 0.5*mu*(Ic - 2*(F.transpose()*R).trace() + 3) + 
           0.5*lambda*(detF - 1).powN(2);
     }
     
     // Eq. 50 from [Smith et al. 2018]
     else if (model == ConstitutiveModels::SaintVenant) {
        Matrix I = energy.make_identity_matrix(3);
        Matrix E = 0.5*(F.transpose()*F - I);
        return mu*E.frobenius_norm_sq() + 
           0.5*lambda*E.trace().powN(2);
     }
  }
\end{mycodeblockscriptsize}

%\input{sections/appendix_manual_differentiation.tex}

%\clearpage
%\input{images/listings.tex}

\end{document}